\documentclass[twocolumn]{aastex7}

\usepackage{amsmath}
\usepackage{amssymb}
\usepackage{graphicx}

\AtBeginDocument{

  \setcounter{dbltopnumber}{2}
  \setcounter{topnumber}{3}
  \setcounter{totalnumber}{4}
}
\usepackage{natbib}
\newcommand{\Msun}{\mathrm{M}_\odot}

\newcommand{\Mpch}{\mathrm{Mpc}\,h^{-1}}

\newcommand{\gadget}{\texttt{GADGET4-Osaka}}
\newcommand{\NSBrunsGNN}{201}

\shorttitle{The CAMELS-CROCODILE Simulation Suite}
\shortauthors{Nagamine et al.}

\begin{document}

\title{The CAMELS-CROCODILE Simulation Suite: \\A New Cosmology--Astrophysics Playground for Machine Learning}

\author[0000-0001-7457-8487,gname=Kentaro,sname=Nagamine]{Kentaro Nagamine}
\affiliation{Theoretical Astrophysics, Department of Earth \& Space Science, Graduate School of Science, The University of Osaka, 1-1 Machikaneyama, Toyonaka, Osaka 560-0043, Japan}
\affiliation{Theoretical Joint Research, Forefront Research Center, Graduate School of Science, The University of Osaka, 1-1 Machikaneyama, Toyonaka, Osaka 560-0043, Japan}
\affiliation{Kavli IPMU (WPI), UTIAS, The University of Tokyo, Kashiwa, Chiba 277-8583, Japan}
\affiliation{Department of Physics \& Astronomy, University of Nevada, Las Vegas, 4505 S. Maryland Pkwy, Las Vegas, NV 89154-4002, USA}
\affiliation{Nevada Center for Astrophysics, University of Nevada, Las Vegas, 4505 S. Maryland Pkwy, Las Vegas, NV 89154-4002, USA}
\email[show]{kn@astro-osaka.jp}

\author[0000-0002-5712-6865,gname=Yuri,sname=Oku]{Yuri Oku}
\affiliation{Theoretical Astrophysics, Department of Earth \& Space Science, Graduate School of Science, The University of Osaka, 1-1 Machikaneyama, Toyonaka, Osaka 560-0043, Japan}
\email{oku@astro-osaka.jp}

\author[0000-0002-6109-2397,gname=Atsushi~J.,sname=Nishizawa]{Atsushi J. Nishizawa}
\affiliation{Institute for Advanced Research, Nagoya University, Nagoya 464-8601, Japan}
\affiliation{Division of Physics and Astrophysical Science, Graduate School of Science, Nagoya University, Nagoya 464-8602, Japan}
\email{atsushi.nishizawa@iar.nagoya-u.ac.jp}

\author[0009-0006-4981-0604]{Jun-Young Lee}
\affiliation{Department of Astrophysical Sciences, Princeton University, 4 Ivy Lane, Princeton, NJ 08544, USA}
\email[]{junyoung.lee@princeton.edu}

\author[0000-0002-4816-0455]{Francisco Villaescusa-Navarro}
\affiliation{Center for Computational Astrophysics, Flatiron Institute, 162 5th Avenue, New York, NY 10010, USA}
\email[]{fvillaescusa@flatironinstitute.org}

\author[0000-0002-3185-1540]{Shy Genel}
\affiliation{Center for Computational Astrophysics, Flatiron Institute, 162 5th Avenue, New York, NY 10010, USA}
\affiliation{Columbia Astrophysics Laboratory, Columbia University, 550 West 120th Street, New York, NY 10027, USA}
\email[]{sgenel@flatironinstitute.org}

\author[0000-0001-5769-4945]{Daniel Angl\'es-Alc\'azar}
\affiliation{Department of Physics, University of Connecticut, 196 Auditorium Road, U-3046, Storrs, CT 06269-3046, USA}
\email[]{angles-alcazar@uconn.edu}

\begin{abstract}
We present CAMELS-CROCODILE, a suite of cosmological hydrodynamic simulations that extends the CAMELS framework with the \gadget{} smoothed particle hydrodynamics code and the Osaka feedback model. As a companion to the IllustrisTNG-based second-generation CAMELS suite, it provides an independent implementation of the baryonic physics that shapes galaxies, the intergalactic medium, and large-scale structure, broadening the range of subgrid models over which machine-learning (ML) inference can be marginalized and tested. The suite comprises $25$ and $50~\Mpch$ boxes organized into one-parameter, cosmic-variance, and Sobol-sequence sets that vary $26$ cosmological and astrophysical parameters around a fiducial Osaka model. We describe the simulation design and validate the suite against benchmarks and other CAMELS suites. Compared with IllustrisTNG, the fiducial model suppresses the matter power spectrum by only a few per cent, against up to ${\sim}27\%$ at $z=0$, and forms stars with similar efficiency below $10^{12.2}~\Msun/h$ but up to $2.5$ times more efficiently in group-scale halos, both consistent with weaker AGN feedback. We also present the stellar and black hole bias and the Ly$\alpha$ absorption around galaxies. As a first ML application, we apply a graph neural network trained on IllustrisTNG galaxy catalogs, with frozen weights, to CAMELS-CROCODILE. It fails to recover $\Omega_{\rm m}$ and $\sigma_8$ and returns overconfident posteriors, mainly for simulations containing more galaxies than any in its training set; within that range $\Omega_{\rm m}$ is recovered with an error only $1.5$ times the in-distribution value. These results underline the need to train and test ML inference across physically distinct galaxy formation models.

\end{abstract}


\section{Introduction}
\label{sec:intro}

The standard $\Lambda$CDM model \citep{Blumenthal84,Hinshaw13} has proven remarkably successful at describing the Universe on large, weakly non-linear scales, where observations of the cosmic microwave background and the clustering of galaxies have constrained the key cosmological parameters to the few-percent level \citep{planck2020}. 
However,  our incomplete understanding of baryonic physics is prohibiting the extraction of comparable information from the smaller, non-linear scales that next-generation surveys probe.  
The cosmic distribution of baryons is governed by the interplay between gravitational collapse and feedback from supernovae (SNe), active galactic nuclei (AGN), and cosmic reionization, all of which redistribute gas and modify the matter power spectrum on scales $k \gtrsim 0.4~h\,\mathrm{Mpc}^{-1}$ in ways that depend strongly on the adopted subgrid model \citep{vanDaalen20,Delgado23,Gebhardt24}. 
Across a library of simulations spanning different galaxy formation implementations, \citet{vanDaalen20} found that the amplitude and scale of this suppression vary widely from model to model while correlating with the baryon fraction of massive halos. 
Varying the feedback strengths systematically instead, \citet{Delgado23} showed that the suppression is controlled by the feedback parameters themselves: stronger AGN feedback lowering halo baryon fractions and deepening the suppression, while stronger stellar feedback can weaken it by stunting black-hole growth. \citet{Gebhardt24} traced the underlying redistribution directly, finding that the fraction of baryons displaced more than $1~$Mpc from their initial dark-matter neighbors ranges from ${\sim}10$ per cent in IllustrisTNG and ASTRID to ${\sim}40$ per cent in SIMBA. Accurately modeling these processes, and crucially, marginalizing over their uncertainty, is therefore essential for interpreting observations of galaxies, the intergalactic medium (IGM), and large-scale structure.

The CAMELS project (Cosmology and Astrophysics with MachinE Learning Simulations) was conceived to address this challenge by reframing it as a machine-learning (ML) problem \citep{villaescusa-navarro2021camels}. Rather than running a handful of expensive flagship simulations, CAMELS comprises thousands of medium-resolution, small-volume simulations that densely sample a joint space of cosmological and astrophysical parameters, providing the large and diverse training sets that neural networks require. This approach has enabled a wide range of ML applications, from inferring $\Omega_{\rm m}$ from two-dimensional density maps to field-level likelihood-free inference from galaxy catalogs, and has repeatedly demonstrated that the dominant obstacle to robust inference is the variation between distinct subgrid physics implementations \citep{villaescusa-navarro2022multifield,ni2023camels,shao2023halos,desanti2023galaxies}. The original CAMELS suite was built around two codes employing the IllustrisTNG and SIMBA galaxy formation models, and was later extended with the ASTRID model and with broader 28-parameter explorations of the TNG and SIMBA spaces \citep{ni2023camels}, precisely because robustness across implementations can only be established by sampling a sufficiently diverse model space. The public release of these datasets has made CAMELS a community-wide resource \citep{villaescusa-navarro2023datarelease}.

Two limitations motivate the present work. First, the CAMELS suites span a range of hydrodynamic codes and galaxy formation models: the moving-mesh code \textsc{Arepo} for TNG, the meshless-finite-mass code \textsc{Gizmo} for SIMBA, and the pressure-entropy SPH code \textsc{MP-Gadget} for ASTRID, the latter two descending from \textsc{GADGET-3}. In these three, the feedback prescriptions are nevertheless structurally similar, in that galactic-wind properties are anchored to galaxy- or halo-scale quantities. A second commonality is calibration: across these models the feedback efficiencies are tuned to reproduce a similar set of low-redshift observables, as in EAGLE \citep{Schaye15}, whose stochastic thermal SN feedback and black-hole accretion and AGN feedback efficiencies are calibrated to the observed $z=0$ galaxy stellar mass function, galaxy sizes, and black-hole--stellar-mass relation \citep{Crain15}. The range of \emph{physically distinct} feedback prescriptions sampled need not, therefore, be as broad as the range of hydrodynamic solvers might suggest. 

It is already clear from within CAMELS that the subgrid choices matter for the machine-learning applications: using synthetic photometry from the IllustrisTNG, SIMBA, ASTRID, and EAGLE suites, \citet{Lovell25} find that a simulation-based inference model trained on one galaxy formation model generalizes poorly to another, a limitation they attribute to differences in the subgrid prescriptions. Second, the canonical CAMELS volume of $(25~\Mpch)^3$ misses long-wavelength modes and rare environments, and suffers from substantial sample variance. The companion second-generation CAMELS suite addresses the latter by increasing the volume to $(50~\Mpch)^3$ while varying 35 parameters of the IllustrisTNG model \citep{genel2026camels50}. The CAMELS-CROCODILE project, presented here, addresses the former: it introduces an independent baryonic physics implementation, the Osaka feedback model, in the smoothed particle hydrodynamics (SPH) code \gadget{}, in which feedback is calibrated to \emph{local} resolved-superbubble physics rather than to galaxy-scale properties. The host code, \textsc{GADGET-4} \citep{Springel21}, brings substantial improvements in force accuracy, time integration, and parallel scalability over its direct predecessor \textsc{GADGET-3} (Section~\ref{sec:code}), and contributes an independently developed, density-independent pressure--energy SPH implementation \citep{Hopk13b} to the methodological diversity already present in CAMELS.

The Osaka model has been developed over a series of papers with the explicit goal of building physically motivated, locally adaptive feedback rather than feedback tuned to halo-scale properties. \citet{shimizu2019osaka} introduced mechanical SN feedback with momentum injection based on Sedov--Taylor-like superbubble solutions in isolated-galaxy tests, while \citet{oku2022osaka} calibrated the energy and momentum loading against resolved Athena++ superbubble simulations and implemented a stochastic thermal plus mechanical scheme in \texttt{GADGET3-Osaka}. These ingredients were brought together in the cosmological CROCODILE simulations \citep{oku2024crocodile}, run with \gadget{}, which add a metallicity- and redshift-dependent top-heavy initial mass function (IMF) for star formation and a black-hole/AGN feedback model that drives metal-rich outflows enriching the circumgalactic and intergalactic medium over Mpc scales. The model has since been validated within the AGORA code-comparison framework \citep{granizo2025agora}, where it has been benchmarked against seven other state-of-the-art codes on the disk formation of Milky Way--mass progenitors at $1 < z < 5$ \citep{Jung25} and on the formation and evolution of galaxies at the high-redshift frontier \citep{Kim26}, and applied to science cases ranging from the cosmic baryon distribution probed by fast radio bursts \citep{crocodile2025frb} to the formation and kinematics of field dwarf galaxies in high-resolution zoom-in simulations \citep{Tomaru26}, demonstrating that the same feedback model operates consistently from dwarf to cosmological scales. Because the Osaka model regulates star formation through a different balance of thermal and mechanical channels than the TNG, SIMBA, or ASTRID models, it samples a genuinely different region of feedback behavior, which is exactly the kind of diversity that robust ML inference demands \citep{villaescusa-navarro2022multifield,ni2023camels}.

In this paper we introduce the CAMELS-CROCODILE suite, embed the Osaka model in the CAMELS parameter-sampling methodology, and present the first global validation of the resulting simulations. 
Section~\ref{sec:overview} describes the simulation design: the code, the two box sizes, the CV, 1P, and SB sets, and the varied parameters. Section~\ref{sec:validation} presents the validation: the halo mass function, the diversity of cosmic star formation histories across the suite, the power spectra of the matter components and the large-scale bias of stars and black holes, the stellar-to-halo mass relation, and the Lyman-$\alpha$ absorption profile around galaxies (Section~\ref{sec:lya_decrement}), compared throughout with established benchmarks and with other CAMELS suites. Alongside the hydrodynamic runs, we also carry out matched dark-matter-only (DMO) counterparts that share the initial-condition phases, which enable clean measurements of the baryonic suppression of the matter power spectrum (Section~\ref{sec:pk_suppression}). Section~\ref{sec:gnn} then gives a first application of the suite as an out-of-distribution test set: a graph neural network trained on the IllustrisTNG-based SB35 set is applied to CAMELS-CROCODILE with frozen weights, testing whether graph-based cosmological inference survives a change of subgrid implementation. Section~\ref{sec:outlook} outlines the planned data release and the ML and large-scale-structure science the dataset is designed to enable.

\section{Simulation Suite Overview}
\label{sec:overview}

\subsection{Code and physics modules}
\label{sec:code}

The CAMELS-CROCODILE simulations are evolved with \gadget{}, the SPH code used for the production CROCODILE runs \citep{oku2024crocodile}, built on \textsc{GADGET-4} \citep{Springel21}. Relative to its widely used predecessor \textsc{GADGET-3}, \textsc{GADGET-4} offers substantially improved gravitational force accuracy, hierarchical time integration that ensures momentum conservation across timestep boundaries, improved parallel scalability through mixed MPI/shared-memory parallelism, and built-in Friends-of-Friends and \textsc{Subfind} group finding. 

Our code solves self-gravity with a TreePM scheme, using a $1024^3$ particle-mesh grid for the long-range force and a tree carrying multipoles to quadrupole order for the short-range force, and hydrodynamics with the density-independent, pressure--energy formulation of SPH \citep{Hopk13b}, in which each particle carries an explicit internal energy variable; this choice (rather than the pressure--entropy variant) allows subgrid feedback energy to be injected directly without a costly iterative treatment \citep{oku2024crocodile}. The SPH implementation employs the Wendland C4 kernel, artificial viscosity based on linearly reconstructed mid-point velocities, artificial conduction with a limiter, and a signal-velocity wake-up timestep limiter \citep[see Appendix A of][]{oku2024crocodile}. 

The code further includes radiative cooling and heating in the presence of a spatially uniform, time-dependent ultraviolet/X-ray background \citep{Haardt12}, star formation in the multiphase interstellar medium, and time-resolved chemical enrichment from Type II and Type Ia SNe and asymptotic giant branch stars via the \texttt{CELib} chemistry library \citep{saitoh2017celib}. Following \citet{oku2024crocodile}, self-shielding of the UV background by neutral hydrogen is not included, to avoid overcooling of the neutral gas. The runs additionally follow the production and destruction of dust with the full grain-size distribution, resolved into $30$ size bins \citep{Aoyama18,Aoyama19,Aoyama20,Romano22a}, and diffuse metals and dust between particles with a Smagorinsky--Lilly turbulent diffusion model \citep[$C_{\rm diff} = 2\times10^{-4}$;][]{Romano22a}.

Stellar feedback follows the Osaka feedback model II \citep{oku2022osaka}: SN energy is deposited through a combination of a stochastic thermal channel, which minimizes spurious radiative losses, and a mechanical (momentum) channel calibrated to resolved superbubble simulations, with the wind velocity and mass loading determined by the \emph{local} density and temperature of the gas rather than by global galaxy or halo properties. 
In the mechanical channel, the neighboring gas particles are projected onto a sphere at the shock radius and partitioned by a Voronoi tessellation on that sphere; the superbubble momentum, mass, and metals are then deposited in proportion to the solid angle $\Omega_i/4\pi$ subtended by each particle's Voronoi face, with the momentum directed along the face normal and an explicit correction term enforcing exact linear-momentum conservation. This solid-angle weighting guarantees isotropic injection even for strongly anisotropic particle distributions, and the same weights are used to distribute the metals from Type II and Type Ia SNe and AGB stars. The energy injection rate per unit stellar mass formed additionally reflects a metallicity- and redshift-dependent top-heavy IMF, which raises the high-redshift feedback budget by up to an order of magnitude relative to a universal IMF \citep{oku2024crocodile}. 

Supermassive black holes are seeded in sufficiently massive halos ($M_h \ge 10^{10}h^{-1}\Msun$) and grow through Bondi rate considering the angular momentum and the viscosity of the
accretion disk on a subgrid scale \citep{rosas-guevara2015accretion}.  AGN feedback couples a fraction of the accretion luminosity to the surrounding gas via stochastic thermal heating, which can naturally produce bipolar outflows (due to the existing galactic disk) that enrich the CGM and IGM.
Analysis of BH growth in CROCODILE is currently underway (Nishihama et al. 2026, in prep).

\subsection{Volumes and resolution}
\label{sec:volumes}

The suite consists of two box sizes, chosen to mirror the first- and second-generation CAMELS volumes and so to enable direct cross-comparisons:
\begin{enumerate}
    \item An L25 set: a $25~\Mpch$ box with $2 \times 256^3$ particles, matching the original CAMELS resolution and volume.
    \item An L50 set: a $50~\Mpch$ box with $2 \times 512^3$ particles, preserving the L25 mass resolution while capturing larger-scale modes, more massive halos, and more diverse environments, and, as in \citet{genel2026camels50}, reducing sample variance.
\end{enumerate}
Because the two boxes share the same mean interparticle spacing, they also share their mass resolution: at the fiducial cosmology the dark-matter particle mass is $m_{\rm DM} = 6.49\times10^{7}~\Msun/h$ and the initial gas particle mass is $m_{\rm gas} = 1.27\times10^{7}~\Msun/h$, scaling with $\Omega_{\rm m}-\Omega_{\rm b}$ and $\Omega_{\rm b}$ respectively across the suite; in the dark-matter-only runs a single particle species carries the total matter mass, $7.75\times10^{7}~\Msun/h$. The gravitational softening is $3.38~{\rm kpc}/h$ comoving, capped at $0.5~{\rm kpc}/h$ physical. Initial conditions are generated at $z = 127$ with second-order Lagrangian perturbation theory (\textsc{2lptic}), and $91$ snapshots are stored per run.
Halos and subhalos are identified with the Friends-of-Friends and \textsc{Subfind} algorithms, and the snapshot, group-catalog, and merger-tree formats follow the CAMELS-IllustrisTNG conventions to maximize interoperability with existing CAMELS analysis tools.

\subsection{Parameter sampling and simulation sets}
\label{sec:params}

The CAMELS-CROCODILE suite varies 26 parameters jointly: 5 cosmological and 21 astrophysical parameters governing star formation, stellar feedback, and AGN/BH physics in the Osaka model \citep{oku2022osaka,oku2024crocodile}. All parameters with their fiducial, minimum, and maximum values are listed in Table~\ref{tab:params}. The four CAMELS-standard labels ($A_{\rm SN1}$, $A_{\rm SN2}$, $A_{\rm AGN1}$, and $A_{\rm AGN2}$) map onto the SN momentum boost, wind mass-loading boost, Bondi accretion boost, and AGN temperature-jump parameters respectively, enabling direct cross-suite comparisons with the TNG- and SIMBA-based CAMELS runs \citep{villaescusa-navarro2021camels,ni2023camels,genel2026camels50}. The remaining 17 astrophysical parameters control finer aspects of the Osaka feedback physics unique to \gadget{}.

The simulations are organized into three sets that share the CAMELS definitions and differ in how the parameter space and the initial-condition random seed are sampled:
\begin{description}
    \item[CV (cosmic variance)] Simulations that all use the fiducial parameter values but differ in their initial-condition random seed, isolating the scatter due to sample variance. Each box size has 27 CV simulations.
    \item[1P (one-parameter)] Variations of a single parameter at a time around the fiducial Osaka model, with all other parameters and the random seed held fixed. This set isolates the effect of each parameter on the simulation outputs and supports clean, like-for-like comparisons. Each parameter is varied at four values, except $n_{\rm spawn}$ and $n_{\rm event,SNII}$, whose integer range $1$--$4$ is covered in full by three values besides the fiducial, and $Z_{\rm HN}$, which is sampled at two; the L25 and L50 1P sets therefore each contain $1 + 100 = 101$ simulations.
    \item[SB (Sobol)] A quasi-random Sobol-sequence sampling of the full 26-dimensional parameter space, providing the large, diverse training set required for ML. The L25 and L50 SB sets contain 128 and 201 (to be expanded to 256) simulations, respectively. Beyond ML training, a Sobol design also supports formal variance-based (Sobol) sensitivity analysis, which measures what fraction of the scatter in a given observable is caused by each parameter separately and what fraction by their joint interactions, typically through an emulator trained on the set. 
    \item[DMO (dark-matter-only)] Matched $N$-body counterparts for every hydrodynamic run, sharing the same initial-condition amplitudes and phases but evolved without baryons. Because the DMO outputs depend only on the cosmological parameters and the random seed, a single fiducial DMO run covers all astrophysics-only 1P variations; separate DMO runs are required only for the cosmological-parameter 1P variations, the CV seeds, and the full SB set. These paired runs enable clean measurements of the baryonic suppression of the matter power spectrum across the parameter space (Section~\ref{sec:pk_suppression}). 
\end{description}

\begin{deluxetable*}{llcrrrl}
\tablecaption{CAMELS-CROCODILE varied parameters.\label{tab:params}}
\tablehead{
  \colhead{CAMELS} & \colhead{Symbol} & \colhead{Description} &
  \colhead{Fiducial} & \colhead{Min} & \colhead{Max} & \colhead{Sampling}
}
\startdata
\multicolumn{7}{c}{\textit{Cosmological (5)}} \\
\hline
\nodata & $\Omega_{\rm m}$   & Matter density parameter              & 0.30  & 0.10  & 0.50  & Linear \\
\nodata & $\sigma_8$         & Amplitude of matter fluctuations      & 0.80  & 0.60  & 1.00  & Linear \\
\nodata & $\Omega_{\rm b}$   & Baryon density parameter              & 0.049 & 0.029 & 0.069 & Linear \\
\nodata & $h$                & Hubble parameter                      & 0.671 & 0.471 & 0.871 & Linear \\
\nodata & $n_s$              & Primordial spectral index             & 0.962 & 0.762 & 1.162 & Linear \\
\hline
\multicolumn{7}{c}{\textit{Star formation (3)}} \\
\hline
\nodata & $\epsilon_*$                  & Star formation efficiency        & 0.01  & 0.002 & 0.05  & Log \\
\nodata & $n_{\rm thres}$ (cm$^{-3}$)  & SF density threshold             & 0.1   & 0.01  & 1.0   & Log \\
\nodata & $n_{\rm spawn}$               & Stars spawned per gas particle   & 2     & 1     & 4     & Linear \\
\hline
\multicolumn{7}{c}{\textit{Stellar/SN feedback (10)}} \\
\hline
$A_{\rm SN1}$ & $f_{p,\rm SN}$    & SN momentum boost factor                  & 1.0  & 0.25 & 4.0  & Log \\
$A_{\rm SN2}$ & $f_{\eta,w}$      & Wind mass-loading boost factor             & 1.0  & 0.25 & 4.0  & Log \\
\nodata & $f_{E,\rm SNII}$        & Type~II SN energy boost factor             & 1.0  & 0.25 & 4.0  & Log \\
\nodata & $f_{E,\rm SNIa}$        & Type~Ia SN energy boost factor             & 1.0  & 0.25 & 4.0  & Log \\
\nodata & $f_{\rm HN}$            & HN fraction at $Z < Z_{\rm HN}$            & 0.5  & 0.1  & 0.9  & Linear \\
\nodata & $Z_{\rm HN}$            & Metallicity threshold for $f_{\rm HN}$      & $10^{-3}$ & $10^{-4}$ & $10^{-2}$ & Log \\
\nodata & $n_{\rm event,SNII}$    & SNII events per star particle              & 2    & 1    & 4    & Linear \\
\nodata & $N_{\rm ngb,fb}$        & Neighbor count for feedback                & 8    & 4    & 12   & Linear \\
\nodata & $f_{c_s,w}$             & Wind temperature (sound-speed) boost       & 1.0  & 0.5  & 2.0  & Log \\
\nodata & $f_{v_w}$               & Wind velocity boost factor                 & 1.0  & 0.5  & 2.0  & Log \\
\hline
\multicolumn{7}{c}{\textit{AGN/BH physics (8)}} \\
\hline
$A_{\rm AGN1}$ & $f_{\rm Bondi}$       & Bondi accretion boost factor        & 1.0  & 0.25 & 4.0  & Log \\
$A_{\rm AGN2}$ & $\Delta T_{\rm AGN}$ (K) & AGN feedback temperature jump    & $3.16\times10^8$ & $7.9\times10^7$ & $1.26\times10^9$ & Log \\
\nodata & $f_{\rm Edd}$           & Eddington accretion boost factor           & 1.0  & 0.1  & 10   & Log \\
\nodata & $C_{\rm visc}$          & BH accretion disc viscosity parameter      & $2\pi\times10^2$ & $2\pi$ & $2\pi\times10^4$ & Log \\
\nodata & $\epsilon_r$            & BH radiative efficiency                    & 0.10 & 0.025 & 0.40 & Log \\
\nodata & $\epsilon_{\rm FB}$     & AGN feedback efficiency                    & 0.15 & 0.0375 & 0.60 & Log \\
\nodata & $M_{\rm seed}$\tablenotemark{a}      & BH seed mass                & $10^{-5}$ & $10^{-6}$ & $10^{-4}$ & Log \\
\nodata & $M_{\rm FoF,seed}$\tablenotemark{a}  & Min.\ FoF mass for BH seeding & 1.0 & 0.1 & 10 & Log \\
\enddata
\tablenotetext{a}{In units of $10^{10}\,h^{-1}\,\Msun$.}
\tablecomments{The ``CAMELS'' column gives the short label used in cross-suite comparisons \citep{villaescusa-navarro2021camels,ni2023camels}; \nodata\ denotes Osaka/\gadget{}-specific parameters without a CAMELS counterpart. ``Sampling'' indicates whether the parameter is varied linearly or log-uniformly in the SB Sobol set. The ``Fiducial'' column is not an arbitrary midpoint of the [Min, Max] range: it is the single calibrated production parameter set of \citet{oku2024crocodile}. 
This fiducial vector anchors the entire suite: the CV set fixes every parameter at these values and varies only the initial-condition seed, and each 1P run holds all 25 remaining parameters at their fiducial values while sweeping the one parameter listed. The SB set instead Sobol-samples the full 26-dimensional volume and need not pass through the fiducial point at all.}
\end{deluxetable*}

\section{Validation and Summary Statistics}
\label{sec:validation}

\subsection{Halo mass function}
\label{sec:hmf}

\begin{figure}
    \centering
    \includegraphics[width=\columnwidth]{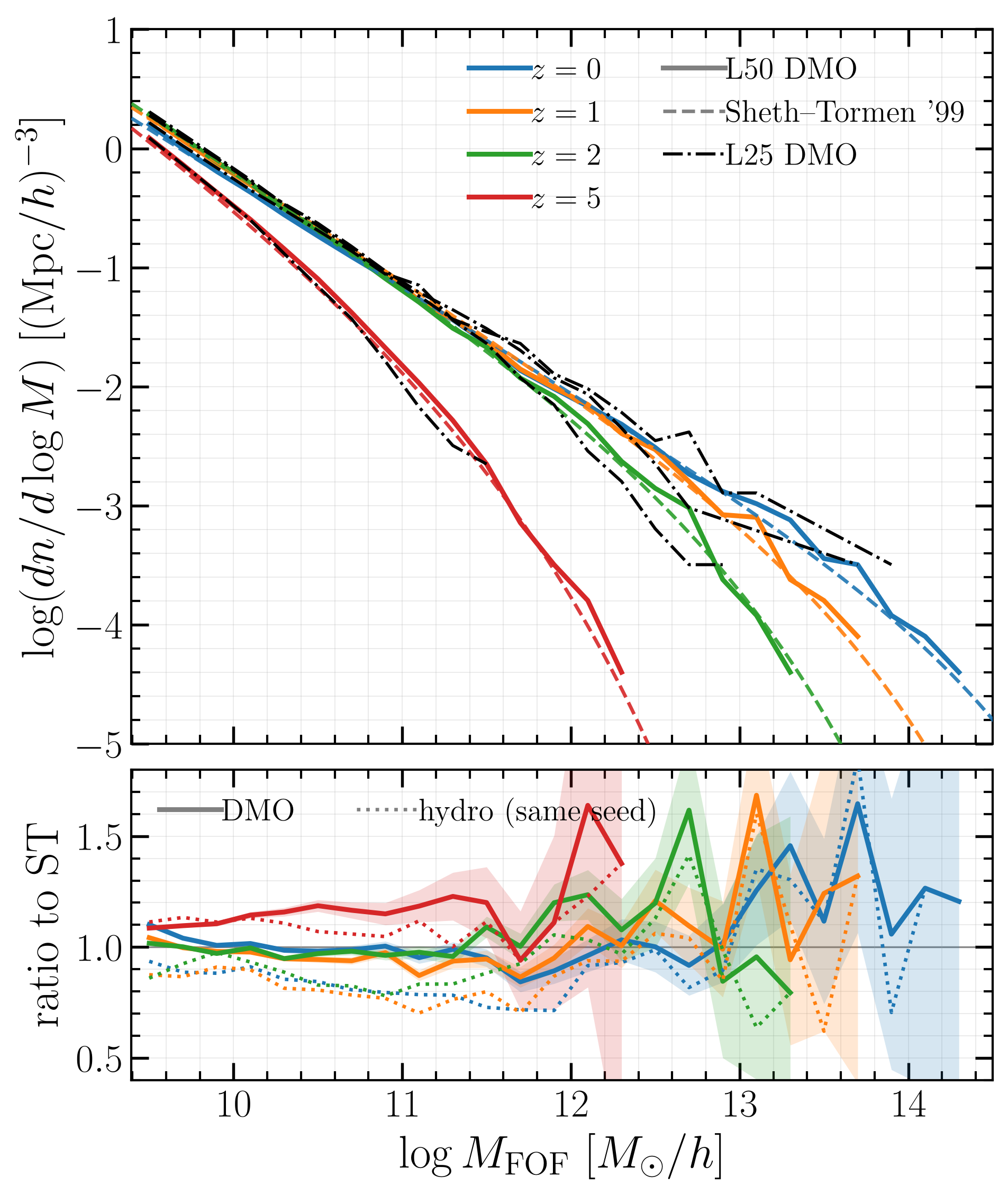}
    \caption{Friends-of-Friends (FOF) halo mass function of CAMELS-CROCODILE. 
    \emph{Top:} the L50 DMO run at $z=0$, $1$, $2$, and $5$ (colored solid curves), the analytic prediction of \citet{Sheth99,Sheth02} evaluated for the CAMELS-CROCODILE fiducial cosmology at the same redshifts (colored dashed curves), and the L25 DMO run (black dash-dotted curves). \emph{Bottom:} ratio of the measured mass function to the Sheth-Tormen prediction, for the L50 DMO run (solid, with the shaded band showing the Poisson uncertainty of the single realization) and for the hydrodynamic L50 run that starts from the \emph{same} initial-condition seed (dotted); the offset between solid and dotted of the same color is therefore the baryonic modification of the halo mass function alone, with no cosmic-variance contribution. All curves start at the FOF completeness floor of $32$ linked particles (Section~\ref{sec:hmf}). Both panels extend to the most massive halo in the box ($1.8\times10^{14}~\Msun/h$ at $z=0$).}
    \label{fig:hmf}
\end{figure}

Before turning to baryonic diagnostics, we first validate the purely gravitational structure formation in the suite, independent of the subgrid feedback physics, by comparing the halo mass function against the analytic prediction of \citet{Sheth99,Sheth02}. Because that prediction is calibrated on $N$-body simulations, we benchmark it against the matched DMO runs (Section~\ref{sec:params}) rather than the hydrodynamic runs, which are shown alongside in the lower panel; the DMO particles carry the total matter mass, so their FOF masses are directly comparable to the theoretical $M_{\rm FOF}$. Figure~\ref{fig:hmf} shows $dn/d\log M$ for FOF halos in the L50 DMO box at $z=0$, $1$, $2$, and $5$, together with the corresponding Sheth-Tormen curves computed for the CAMELS-CROCODILE fiducial cosmological parameters.

The low-mass limit of every curve is set by the group finder itself: \gadget{} retains only FOF groups with at least $32$ linked particles, so the catalogs are complete above $32\,m_{\rm p} = 2.48\times10^{9}~\Msun/h$. Above that floor the agreement with Sheth-Tormen is excellent at $z \le 2$: the ratio has a median of $0.99$, $0.98$, and $1.00$ at $z=0$, $1$, and $2$ over the three decades of mass where the box holds at least $20$ halos per bin, and stays within roughly $5$--$10$ per cent of unity there. Above $\sim\!10^{12.5}~\Msun/h$ the excursions grow to tens of per cent, but they are sampling noise rather than a systematic offset: the same excursions appear in the same bins in the hydrodynamic run, which contains the same halos. This confirms that the initial conditions, linear growth, and gravity solver are all behaving as expected. At $z=5$ the measured mass function lies systematically above the fit, by $16$ per cent in the median; we have not traced the origin of this offset, which does not affect the baryonic comparisons below, as those are ratios taken at fixed initial conditions. 

The lower panel also quantifies how much the baryons move the halo mass function. Since the hydrodynamic run shown there shares its initial-condition seed with the DMO run, the hydro-to-DMO ratio is free of cosmic variance. 
At $z=0$ the hydrodynamic run has $15$ per cent fewer halos than the DMO run at $10^{9.5}~\Msun/h$ and $14$ per cent fewer at $10^{10.5}~\Msun/h$, with the largest deficit, $23$ per cent, at $10^{11.5}~\Msun/h$; the ratio then returns to unity in the group and cluster regime, where the deep potential wells retain their baryons. The same pattern holds at $z=1$ and $z=2$ and weakens toward $z=5$ ($\sim\!6$ per cent in the median). This is the halo-mass-function counterpart of the power-spectrum suppression $S(k)$ analyzed in Section~\ref{sec:pk_suppression}: feedback drives gas out of low- and intermediate-mass halos, lowering their total FOF mass and shifting them down the mass function. We emphasize that $M_{\rm FOF}$ is the total mass of all species in both runs, so this offset is not the trivial $(\Omega_{\rm m}-\Omega_{\rm b})/\Omega_{\rm m} = 0.837$ rescaling that a dark-matter-only mass would produce.
The physical origin is the near-complete evacuation of baryons from low-mass halos: the median halo baryon fraction is only $12$--$18$ per cent of the cosmic value below $10^{11.4}~\Msun/h$, rising to $\sim\!90$ per cent by $10^{13.4}~\Msun/h$. Holding the dark matter fixed, those baryon fractions imply $M_{\rm hydro}/M_{\rm DMO} \simeq 0.86$ at the low-mass end, which, given the local slope $d\log n/d\log M \simeq -1$, accounts for essentially the entire observed deficit there. It also shows why the DMO runs are the right reference for a validation against an $N$-body-calibrated fit, since using the hydrodynamic mass function instead would have introduced a $\sim\!15$ per cent offset of purely baryonic origin.

As a further consistency check, we overlay the L25 DMO run (black dash-dotted curves), which has the same particle mass as L50 but probes an $8\times$ smaller volume. At the low-mass end, where both boxes contain large numbers of halos, the two agree to $3$--$5$ per cent at every redshift. Toward higher masses the single L25 realization departs from L50 by $10$--$25$ per cent, and its curves terminate well before the L50 exponential cutoff. Both effects are sample variance rather than resolution: a $25~\Mpch$ box contains few of the rare massive halos and misses the large-scale modes that seed them, which is precisely why the L50 volume was added to the suite.

\subsection{Cosmic star formation history}
\label{sec:sfrd}

\begin{figure*}
    \centering
    \includegraphics[width=\textwidth]{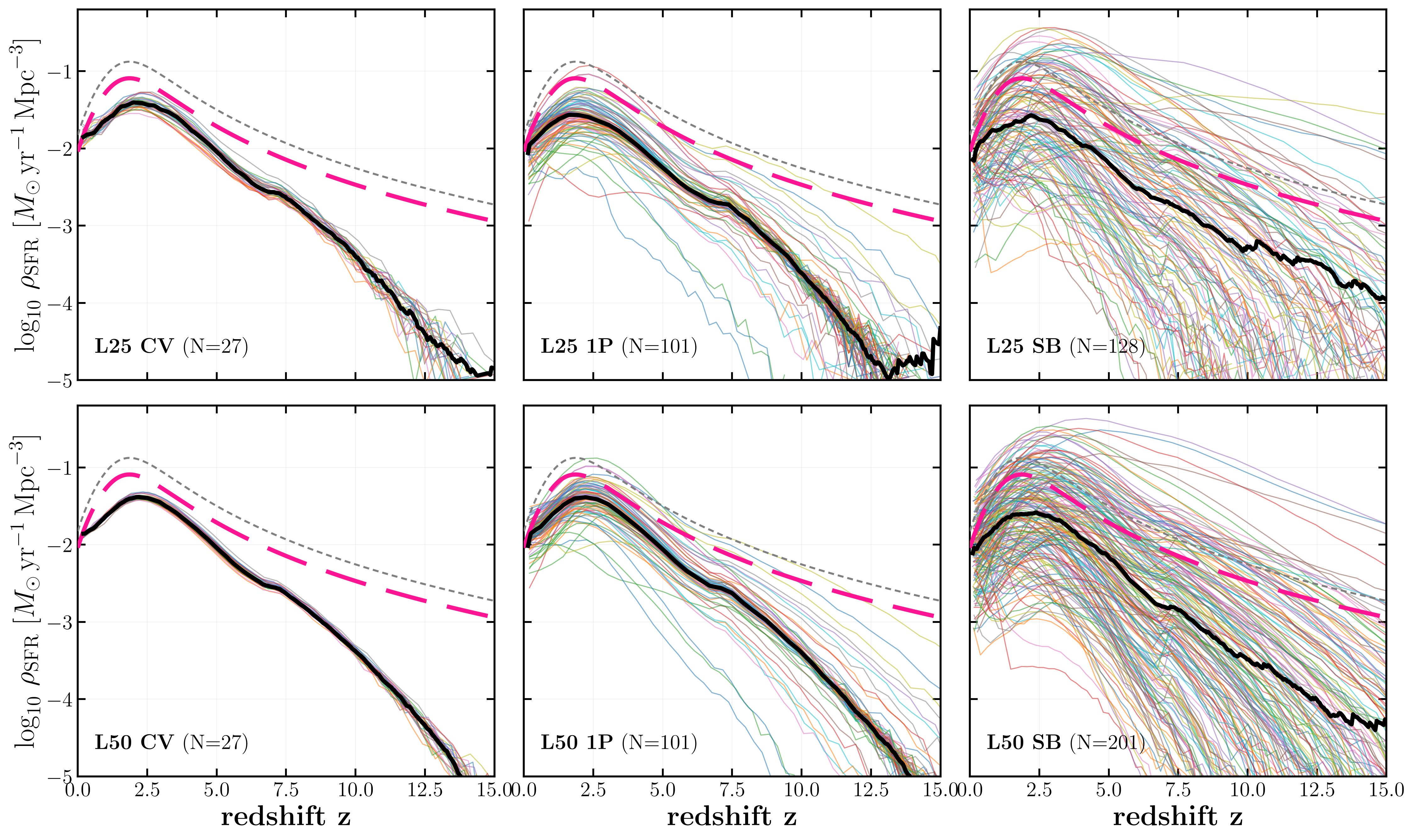}
    \caption{Cosmic star formation rate density $\rho_{\rm SFR}$ as a function of redshift across the CAMELS-CROCODILE suite. The top row shows the L25 runs and the bottom row the L50 runs, for the cosmic-variance (CV), one-parameter (1P), and Sobol (SB) sets from left to right, with the number of simulations in each panel indicated. 
    Thin colored curves show individual simulations and the thick black curve shows the set median. For each realization, we use its own value of $h$ when converting particle masses in $M_\odot/h$ and comoving volumes in $(\mathrm{Mpc}/h)^3$. 
    The 1P and SB sets exhibit a broad spread driven by the varied feedback, whereas the CV set shows a tight locus that quantifies the sample variance, markedly smaller in the larger L50 volume. The gray short-dashed line shows the observational fit of \citet{madau2014review} with Salpeter IMF, and the thick pink long-dashed line for Chabrier IMF (scaled down by a factor of 1.64) for reference.}
    \label{fig:sfrd}
\end{figure*}

Having confirmed that the underlying gravitational structure formation is correct, we turn to the cosmic star formation rate density (SFRD), $\rho_{\rm SFR}(z)$, which integrates the self-regulation of star formation by feedback across all halos and cosmic times and provides the first test of the baryonic and subgrid physics. Figure~\ref{fig:sfrd} presents $\rho_{\rm SFR}(z)$ for the full CAMELS-CROCODILE suite, with the L25 box in the top row and the L50 box in the bottom row, and the CV, 1P, and SB sets shown from left to right.

Three features are immediately apparent. First, the median SFRD in every panel rises from high redshift to a peak at $z \sim 2$--$3$ and declines toward $z=0$, in qualitative agreement with the observed cosmic star formation history \citep[e.g.,][]{madau2014review} and with the self-regulated behavior expected of physically motivated feedback model \citep{springel2003sfr,sorini2021sfr}. 

Second, the 1P and SB sets display a wide diversity of histories, driven by the cosmological parameters as much as by the astrophysical ones: the 1P set sweeps each of the 26 parameters in turn, five of which are cosmological, and SB26 samples all 26 jointly. 
Quantifying the L50 1P spread at fixed redshift by the ratio of the largest to the smallest $\rho_{\rm SFR}$ among the variants of a single parameter, the astrophysical parameters lead below $z \simeq 2$: the star formation efficiency (a factor of $7.4$ at $z=0.5$), the black-hole accretion-disc viscosity ($6.6$ at $z=1$), and the Type~II SN energy ($12.5$ at $z=2$). At higher redshift the cosmological parameters take over. At $z=4$ the two largest are $\Omega_{\rm m}$ ($20.5$) and $\sigma_8$ ($17.5$), with the Type~II SN energy essentially tied for third ($17.4$); by $z=6$ the three largest are all cosmological, $\Omega_{\rm m}$ ($55.9$), $\sigma_8$ ($36.1$) and $n_s$ ($27.8$), against $6.1$ for the strongest astrophysical parameter. 
The total 1P spread is therefore indeed largest at high redshift, but because the cosmological variations change how much structure has collapsed by then, not because feedback couples more efficiently to low-mass halos: the spread across the $80$ astrophysics-only runs alone peaks at $z \simeq 2$ (a factor of $15.5$) and falls to $7.2$ by $z=6$. This diversity, cosmological and astrophysical together, is precisely the property that makes the suite valuable as an ML training set, since it forces inference models to learn mappings robust to both rather than memorizing a single history. 

Third, the CV sets (which fix all parameters and vary only the initial random seed) occupy a narrow band whose width measures the sample variance alone. As expected, the CV scatter is substantially smaller than in the 1P \& SB set, and it is visibly reduced in the larger L50 volume relative to L25, consistent with the volume scaling of cosmic variance characterized in \citet{genel2026camels50}, who find that the CV-set scatter in the cosmic SFRD shrinks by a factor of $\approx 2$ between $25$ and $50~h^{-1}{\rm Mpc}$ boxes, somewhat below the naive $\sqrt{(50/25)^3} \approx 2.8$ scaling expected for independent volumes, an effect they attribute to mode coupling between octants of the larger box. This hierarchy (tight CV scatter versus broad parameter-sampling (1P/SB) scatter) reproduces the same qualitative behavior seen in the summary-statistic comparisons of the original CAMELS suite \citep{villaescusa-navarro2021camels,ni2023camels}, where the CV sets isolate sample variance alone and the Latin-hypercube(LH) sets are deliberately designed to span a much wider range by additionally varying cosmological and astrophysical parameters.

Our \gadget{}-based runs reproduce the cosmic star formation history and yield physically plausible gas and metal distributions, while the parameter-induced diversity spans and in places extends beyond the range covered by the existing CAMELS suites. Because the Osaka model reaches a given level of star formation suppression through a different combination of thermal and mechanical feedback than the TNG, SIMBA, and ASTRID models, the suite provides an independent test of the robustness of ML inference pipelines trained on other subgrid implementations \citep{ni2023camels}.

\subsection{Power spectra of the matter components}
\label{sec:pk_components}

\begin{figure*}
    \centering
    \includegraphics[width=\textwidth]{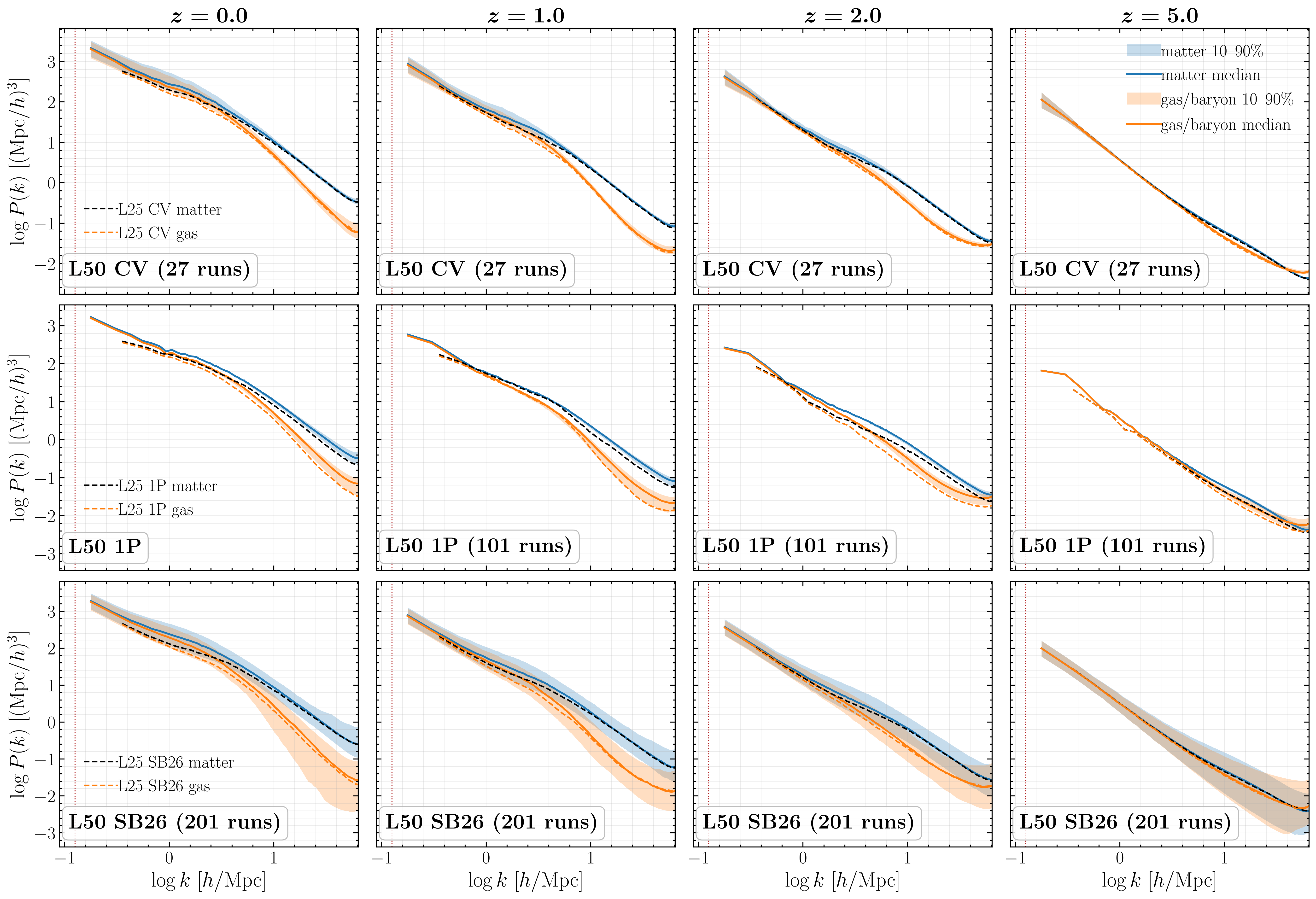}
    \caption{Power spectra of the total matter (blue) and gas (orange) in the L50 simulation sets at $z = 0$, $1$, $2$, and $5$ (left to right columns). The top, middle, and bottom rows show the CV, 1P, and SB26 sets respectively, with the number of contributing runs given in each panel.
    In every panel the thick curve is the ensemble median over the runs of that set and the shaded band their 10--90 percentile range. 
 The gas power spectrum is increasingly suppressed relative to the dark matter toward low redshift on scales $k \gtrsim$ a few $h\,\mathrm{Mpc}^{-1}$, reflecting the combined action of thermal pressure support and feedback-driven redistribution of baryons. 
    The SB26 panels show a substantially wider spread than the CV panels, since the Sobol sampling simultaneously varies all 26 cosmological and astrophysical parameters, while the CV set holds them fixed; the 1P band, which varies one parameter at a time about the fiducial model, is intermediate. 
    Each panel additionally overlays the median matter (black dashed) and gas (orange dashed) power spectra of the corresponding L25 set, which agree closely with the L50 result, except that they start at a higher $k$ because the smaller box has a coarser fundamental mode. 
    The red dotted vertical line marks the box fundamental mode, $k_{\rm f} = 2\pi/L_{\rm box}$ (at $\log k = -0.90$), below which the power spectrum is not sampled, and the right-hand edge of each panel is the Nyquist frequency, $k_{\rm Ny} = \pi N^{1/3}/L_{\rm box}$ (at $\log k = 1.81$).}
    \label{fig:pk_cv}
\end{figure*}

A second basic summary statistic of the suite is the power spectrum of the matter components. Figure~\ref{fig:pk_cv} shows the total matter and gas power spectra of the L50 CV, 1P, and SB26 sets at $z = 0$--$5$. At high redshift the gas closely traces the dark matter, while toward $z = 0$ the gas power is progressively suppressed on scales $k \gtrsim$ a few $h\,\mathrm{Mpc}^{-1}$ as thermal pressure support and feedback redistribute baryons out of halos, the same physical effect that drives the hydro-to-$N$-body suppression $S(k)$ quantified in Section~\ref{sec:pk_suppression}.

The scatter across the three sets reflects distinct sources of variance. In the CV set, which holds cosmology and feedback fixed and varies only the initial-condition phase, the fractional 10--90 percentile spread in the gas power spectrum is largest on large scales (${\sim}90\%$ at $k \approx 0.2~h\,{\rm Mpc}^{-1}$), where a $50~\Mpch$ box samples few independent long-wavelength modes, and comparatively narrower on small scales (${\sim}70\%$ at $k \approx 40~h\,{\rm Mpc}^{-1}$), where each realization averages over many independent halos under identical feedback physics. 
The 1P set shows the opposite pattern: because its parameter variations share fixed initial-condition phases by design, the large-scale scatter is small (${\sim}5\%$), reflecting the weak sensitivity of large-scale power to feedback parameters, while the small-scale scatter is larger than in the CV set (${\sim}80\%$), since several of the varied parameters (Table~\ref{tab:params}) directly control the SN/AGN feedback strength that sets the depth of small-scale baryonic suppression. 
The SB26 set, which simultaneously varies all 26 parameters across the full prior volume, displays the widest spread at all scales, as the joint variation of cosmological and astrophysical parameters compounds both the large-scale cosmic-variance contribution and the small-scale feedback sensitivity. The three band widths therefore encode complementary quantities: sample variance (CV), single-parameter sensitivity (1P), and the full joint parameter response (SB26).

\subsection{Baryonic suppression of matter clustering}
\label{sec:pk_suppression}

\begin{figure*}[!t]
    \centering
    \includegraphics[width=\textwidth]{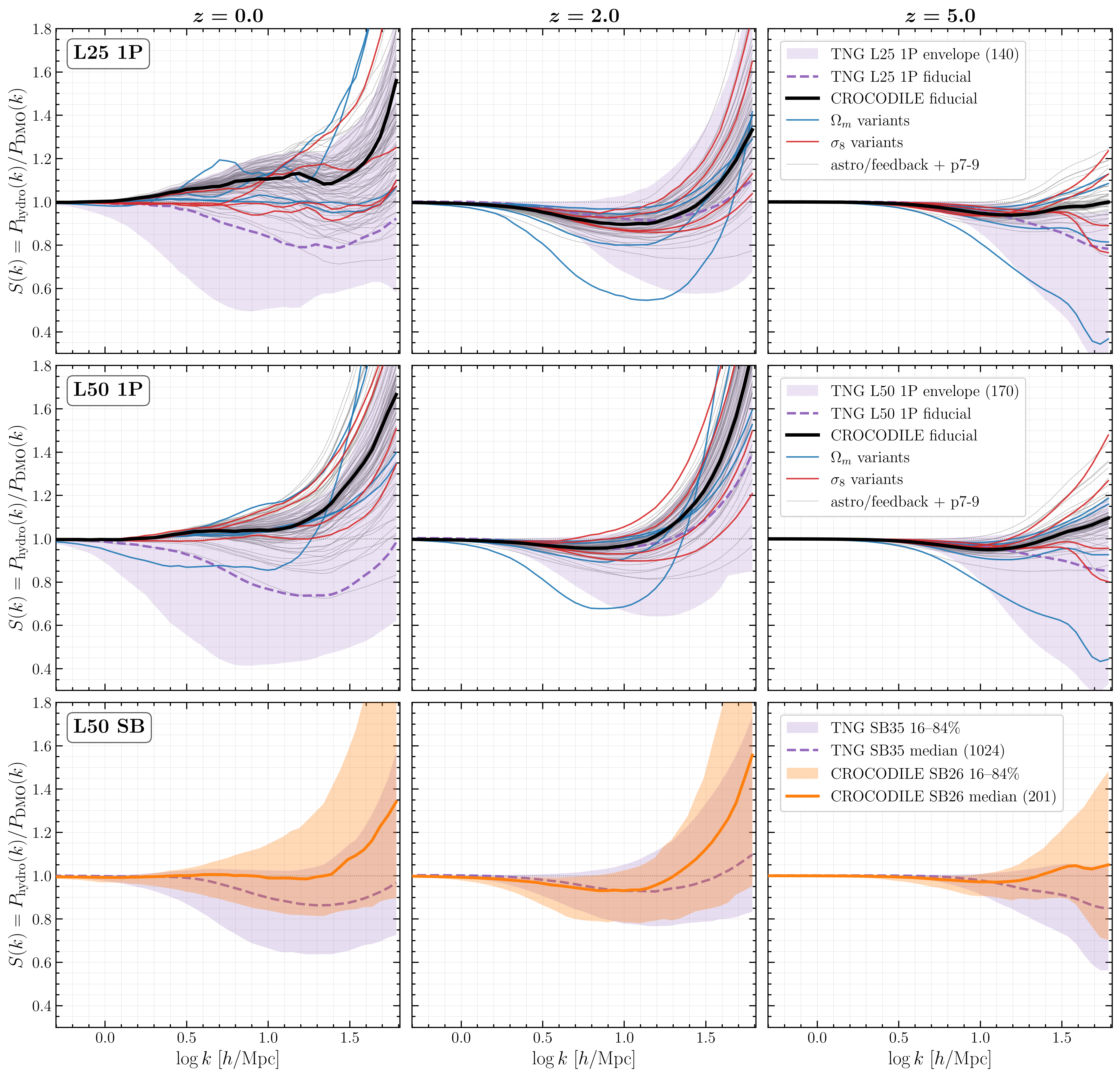}
    \caption{Baryonic suppression of the matter power spectrum, $S(k) = P_{\rm hydro}(k)/P_{\rm DMO}(k)$, at $z = 0$, $2$, and $5$ (left to right), compared with the IllustrisTNG suite in purple.
    \emph{Top row:} the L25 1P set; \emph{middle row:} the L50 1P set.
    In these two rows the thick black curve shows the CROCODILE fiducial model, blue curves trace the four $\Omega_{\rm m}$ variants, red curves the four $\sigma_8$ variants, and thin gray curves all remaining astrophysics-only variations (p3--p6, p10--p26) plus $\Omega_{\rm b}$, $h$, and $n_s$ (p7--p9), which share the fiducial DMO as their denominator; the purple shaded region is the full min--max envelope of the box-matched IllustrisTNG 1P family and the dashed purple curve is its fiducial run. \emph{Bottom row:} the L50 Sobol sets, where both suites are summarized by their ensemble median (solid orange for CROCODILE SB26, dashed purple for IllustrisTNG SB35) and $16$th--$84$th percentile band. 
    }
    \label{fig:pk_suppression}
\end{figure*}

The matched $N$-body counterparts of every hydrodynamic run allow a clean measurement of the baryonic suppression of the total matter power spectrum, $S(k) \equiv P_{\rm hydro}(k)/P_{\rm DMO}(k)$, where $P_{\rm hydro}(k)$ is the total matter spectrum in the hydrodynamic run.\footnote{The power spectra are tabulated on a linear $k$ grid, so on the logarithmic abscissa used here the sampling is heavily weighted toward high $k$; the curves are therefore formed as a mode-weighted band power, averaging $P_{\rm hydro}$ and $P_{\rm DMO}$ separately over $20$ logarithmic $k$ bins per decade (weighting each sample by $k^2$) before taking their ratio. This suppresses the single-realization sampling noise that otherwise dominates the appearance of the curves at $k \gtrsim 10~h\,{\rm Mpc}^{-1}$, and it is applied identically to the CROCODILE and IllustrisTNG curves. At low $k$ the linear sampling is coarser than the logarithmic bins, so below the scale at which the two coincide ($k \simeq 1.1$ and $2.2~h\,{\rm Mpc}^{-1}$ for the $50$ and $25~\Mpch$ boxes, respectively) the native sampling is retained rather than rebinned.} 
Beyond cataloging the suppression itself (which the TNG-, SIMBA-, and ASTRID-based suites have characterized within their respective models \citep{villaescusa-navarro2021camels,ni2023camels,genel2026camels50}, our longer-term goal is to use CAMELS-CROCODILE to test the proposed quasi-universal relation between $S(k)$ and the baryon fraction of group-scale halos \citep{vanDaalen20,Loon24}. 
   That relation has so far been established predominantly with a limited family of subgrid models and hydrodynamic methods. 
   
   The physical origin of the link is well illustrated by \citet{sorini2022baryons}, who compared SIMBA against its dark-matter-only counterpart and traced how feedback redistributes baryons: in that model, star-formation-driven outflows dominate at $z > 2$ in lower-mass halos while AGN jets dominate at $z < 2$ in higher-mass halos, evacuating gas so efficiently that the baryon fraction within $\sim 10^{12}$--$10^{13}~\Msun$ halos falls to only $\sim 25\%$ of the cosmic value, and returning to the cosmic fraction only when integrated out to $\sim 10$--$20$ virial radii. 
  They further reported that this redistribution alters the halo mass function itself at the ${\sim}20\%$ level. These are SIMBA-specific numbers and we do not adopt them as a benchmark, but they make explicit why $S(k)$ and the halo baryon fraction are expected to be coupled, and why the coupling should be probed across independently implemented feedback models. 
  CAMELS-CROCODILE tests its universality with an independent physical implementation: locally calibrated mechanical SN feedback and purely thermal AGN energy injection.
A concrete example of the kind of prior at stake is provided by \citet{sorini2024profiles}, who fitted the halo gas density profiles of six SIMBA feedback variants over $10^{11}~\Msun < M_{\rm 200c} < 10^{14}~\Msun$ and $0 < z < 4$ with a universal power-law formula intended to be applied to $N$-body simulations and semi-analytic models; such formulae inherit whatever feedback model calibrated them.
  As a first step toward that test, we characterize the suppression itself across the L25 \& L50 1P parameter sweep.

Figure~\ref{fig:pk_suppression} shows the resulting $S(k)$ for the full 1P set in both boxes (top and middle rows) and for the L50 Sobol set (bottom row), at $z = 0$, $2$, and $5$. At low $k$ ($\lesssim 1~h\,{\rm Mpc}^{-1}$), all curves converge to $S \approx 1$ regardless of redshift, box size, or parameter choice, as expected: baryonic feedback redistributes mass only on scales comparable to or smaller than individual halos, leaving the large-scale linear power untouched. 
The fiducial model itself shows only shallow suppression at any redshift, dipping no more than a few percent below unity ($S_{\rm min} \sim 0.95$--$0.99$) before rising to an \emph{enhancement} above unity at high $k$; the individual astrophysical parameter variants, however, span a much broader range, with the most extreme few reaching $S \sim 0.7$--$0.9$ at intermediate $k$ ($k \sim 3$--$10~h\,{\rm Mpc}^{-1}$) depending on redshift, illustrating how directly the SN/AGN feedback and star-formation parameters set the depth of the baryonic suppression. We tentatively attribute the high-$k$ enhancement ($\gtrsim 50~h\,{\rm Mpc}^{-1}$) to cooling and star condensation outpacing feedback at small scales.  This regime also coincides with CIC aliasing near the Nyquist frequency ($k_{\rm Ny} \approx 64.3~h\,{\rm Mpc}^{-1}$) and shot noise, but we checked that the small-scale enhancement remains even after recomputing the power spectrum with interlacing. The shallowness of this suppression is not in tension with the pronounced deficit of the gas power spectrum in Figure~\ref{fig:pk_cv}: baryons enter the total matter field with weight $f_{\rm b} \simeq 0.16$, so their own power contributes only as $f_{\rm b}^2 \simeq 0.03$ together with the cross term, and the gas removed from small scales is converted into stars, which are far more strongly clustered than the dark matter and have no counterpart in the $N$-body run. In the fiducial L50 model at $z=0$, for example, $P_{\rm gas}$ falls to $0.48$ and $0.25$ of $P_{\rm m}$ at $k = 10$ and $40~h\,{\rm Mpc}^{-1}$, while the dark matter component alone differs from its DMO counterpart by less than $6\%$ out to $k = 10~h\,{\rm Mpc}^{-1}$; the shallow $S(k)$ therefore reflects a genuinely weak redistribution of mass rather than a cancellation between strongly affected components.

The gray envelope in each panel, spanning the 21 astrophysical parameters (star formation, SN, and AGN feedback) plus the three cosmological shape parameters $\Omega_{\rm b}$, $h$, and $n_s$, has a median close to the shallow fiducial curve at every redshift, but its most extreme individual members reach substantially deeper suppression, particularly for the star-formation and AGN parameters. The two headline cosmological parameters, $\Omega_{\rm m}$ and $\sigma_8$, produce a spread comparable to or exceeding that of the most extreme astrophysical variants. The L50 1P set isolates their effect most cleanly, since each of these variants has its own matched $N$-body counterpart and differs from the fiducial run in one parameter only, and we quote it here. Both parameters act in the same direction: lowering either one deepens the suppression, monotonically across the sampled range. At $z=2$, where the minima lie at $k \simeq 3$--$13~h\,{\rm Mpc}^{-1}$,
$S_{\rm min} = 0.678$, $0.893$, $0.956$, $0.980$, and $0.989$ for $\Omega_{\rm m} = 0.10$, $0.20$, $0.30$ (fiducial), $0.40$, and $0.50$, and $S_{\rm min} = 0.898$, $0.930$, $0.956$, $0.980$, and $0.989$ for $\sigma_8 = 0.60$ to $1.00$ in steps of $0.10$.

At $z=2$ the two parameters therefore appear to act alike, but that agreement is specific to the depth of the dip, and the $z=0$ panel separates them: $\sigma_8$ displaces the whole $S(k)$ curve coherently, whereas $\Omega_{\rm m}$ changes its shape. Lowering $\sigma_8$ lowers $S$ at every scale, so that at $z=0$ and $k = 40~h\,{\rm Mpc}^{-1}$ the enhancement falls monotonically from $S = 1.631$ at $\sigma_8 = 1.00$ to $1.100$ at $\sigma_8 = 0.60$; the dip position moves in step, migrating from $k = 3.2$ to $12.5~h\,{\rm Mpc}^{-1}$ at $z=2$ as $\sigma_8$ falls over the same range, tracking the nonlinear scale to smaller scales. Lowering $\Omega_{\rm m}$ does not act coherently: the $\Omega_{\rm m} = 0.10$ run is simultaneously the lowest curve in the panel at $k = 10~h\,{\rm Mpc}^{-1}$ ($S = 0.856$) and the highest at $k = 40~h\,{\rm Mpc}^{-1}$ ($S = 1.893$), crossing unity between the two, and the $\Omega_{\rm m}$ ordering at high $k$ runs opposite to the $\sigma_8$ ordering. Any statement about the sign of the cosmological dependence of $S(k)$ consequently requires a scale attached to it.

Both behaviors follow from reading $S(k)$ as a competition between two effects of opposite sign: gas that is pressure-supported or driven out of halos is smoother than the dark matter and pushes $S$ below unity, while stars, which are more strongly clustered than any other component and have no counterpart in the $N$-body run, push it above. What governs the stellar term is not the absolute stellar density but the share of the matter budget locked into stars, $f_\star = \rho_\star/(\Omega_{\rm m}\rho_{\rm crit})$, and the 1P set allows the two candidates to be told apart. Along the $\sigma_8$ sequence, where $\Omega_{\rm m}$ is fixed and the two are proportional, $f_\star$ rises from $0.0032$ to $0.0104$ and $S(k = 40~h\,{\rm Mpc}^{-1}, z=0)$ rises monotonically with it. Along the $\Omega_{\rm m}$ sequence they separate: $\rho_\star$ \emph{increases} with $\Omega_{\rm m}$, from $3.1$ to $7.3 \times 10^{8}~\Msun\,h^2\,{\rm Mpc}^{-3}$, and is therefore anticorrelated with $S$, while $f_\star$ \emph{decreases}, from $0.0113$ to $0.0053$, and stays rank-correlated with $S$, as it is along the $\sigma_8$ sequence. The absolute stellar density is thus excluded as the controlling variable, whereas the stellar mass fraction orders both sequences correctly. The low-$\Omega_{\rm m}$ run forms the fewest stars of the family in absolute terms, having the least matter to work with, yet converts the largest share of its matter into them.

That share is set by the baryon supply. Because the 1P design varies one parameter at a time, $\Omega_{\rm b}$ is held fixed as $\Omega_{\rm m}$ moves, so the cosmic baryon fraction $f_{\rm b} = \Omega_{\rm b}/\Omega_{\rm m}$ varies along the $\Omega_{\rm m}$ sequence from $0.098$ at $\Omega_{\rm m} = 0.50$ to $0.49$ at $\Omega_{\rm m} = 0.10$, three times its fiducial value. A larger baryon reservoir amplifies both competing terms at once, deepening the intermediate-$k$ dip and strengthening the small-scale stellar enhancement, which is why $\Omega_{\rm m}$ alters the shape of $S(k)$ where $\sigma_8$, leaving $f_{\rm b}$ untouched, merely displaces it. Consistently, $\Omega_{\rm m}$ is the stronger of the two at $z=2$, spanning a deficit $1 - S_{\rm min}$ of $0.011$--$0.322$ against $0.011$--$0.102$ for $\sigma_8$.

We do not quote the $z=5$ minima for the two most extreme variants: they fall at $k = 54$ and $61~h\,{\rm Mpc}^{-1}$, within the regime where CIC aliasing and shot noise preclude a firm interpretation. This overlap between the astrophysical envelope's extremes and the cosmological spread previews the difficulty of the intended universality test: to isolate the halo-baryon-fraction dependence of $S(k)$ cleanly, the comparison must be performed at fixed cosmology, or the cosmological trend must be modeled and divided out.

The purple curves in Figure~\ref{fig:pk_suppression} place these results alongside the CAMELS-IllustrisTNG suite, matched box-for-box and design-for-design: the CROCODILE 1P sets are compared against the IllustrisTNG 1P sets in the same volume, and the L50 SB26 set against the IllustrisTNG SB35 set. Because both suites adopt the same CAMELS one-parameter-at-a-time construction and the same parameter numbering for the cosmological variations, and because their power spectra are measured on a common $k$ grid at identical mass resolution, the comparison is as direct as possible. The contrast is striking: the IllustrisTNG fiducial model suppresses the matter power spectrum by up to $\sim 27\%$ at $z=0$ ($S_{\rm min} \simeq 0.74$ at $k \sim 20~h\,{\rm Mpc}^{-1}$ in L50, and $S_{\rm min} \simeq 0.78$ in L25), whereas the CROCODILE fiducial remains within a few percent of unity at all $k$ and crosses to enhancement at high $k$. The same offset persists between the Sobol ensembles, where the IllustrisTNG SB35 median reaches $S \simeq 0.86$ at $z=0$ against $S \simeq 0.99$ for CROCODILE SB26, and it holds at every redshift shown, with the two suites agreeing most closely at $z=2$.

We caution that this is a comparison of two fully independent subgrid implementations, not a controlled experiment: the CROCODILE and IllustrisTNG parameter sets are not in one-to-one correspondence beyond the cosmological parameters, so the offset reflects the combined effect of differing feedback channels, hydrodynamic solvers, and calibration targets. Nonetheless, its size and persistence across box sizes, sampling designs, and redshifts indicate that CROCODILE's locally calibrated mechanical SN feedback and thermal AGN injection redistribute substantially less mass out of halos than IllustrisTNG's kinetic wind and radio-mode AGN model. This is precisely the kind of model-to-model spread that makes CAMELS-CROCODILE informative as an independent test of the proposed $S(k)$--$f_{\rm b}$ universality: if the relation is genuinely model-independent, these two very different suppression amplitudes should still map onto a common locus once each is referred to its own halo baryon fraction.

\subsection{Tracer bias: stars and black holes}
\label{sec:xpk}

\begin{figure*}
    \centering
    \includegraphics[width=0.94\textwidth]{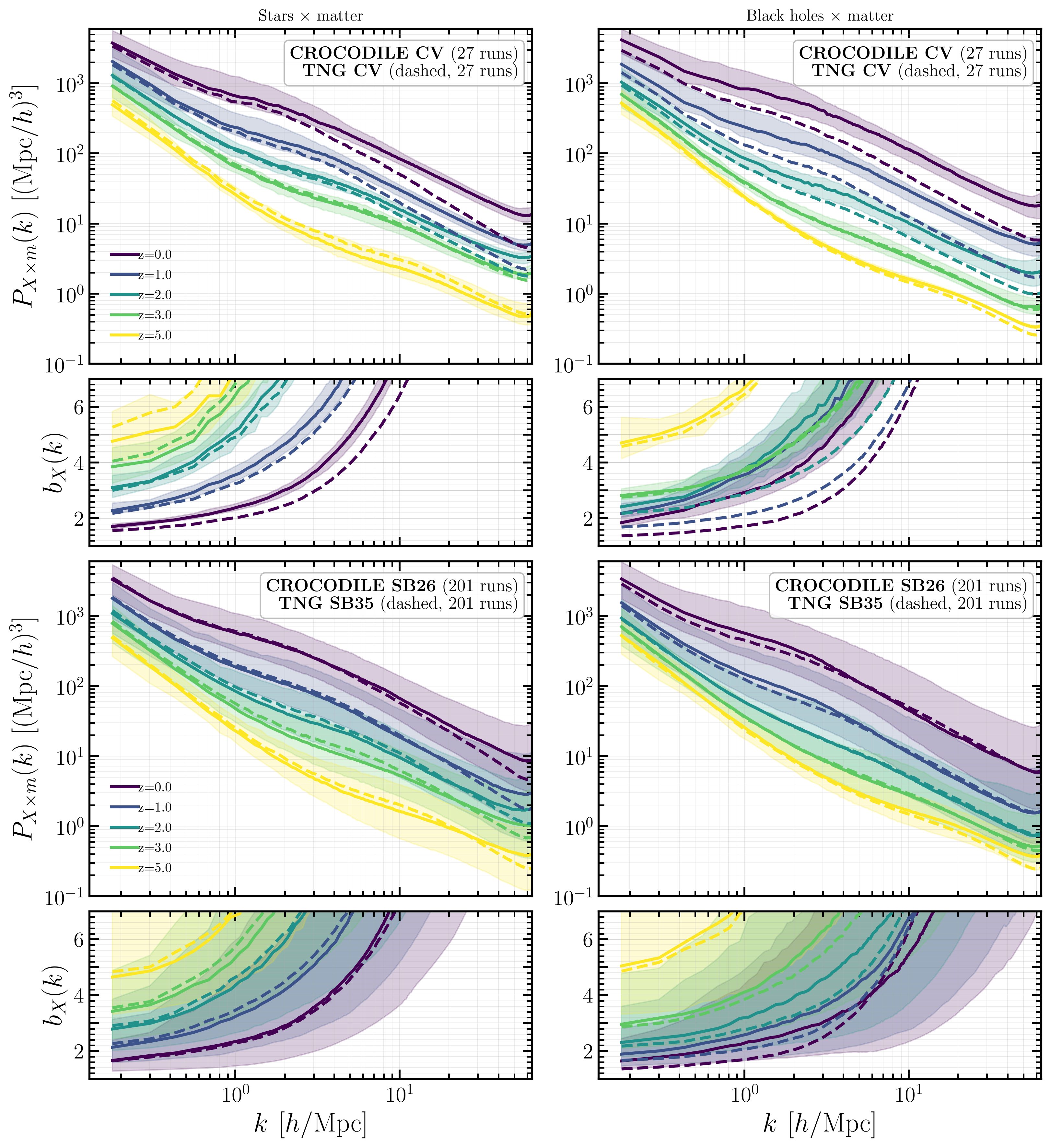}
    \caption{Redshift evolution of the cross-power spectra and implied large-scale bias of stars (left column) and black holes (right column) with the total matter field, for the L50 CV set (27 realizations; top two rows) and the L50 SB26 set (201 realizations; bottom two rows). The first and third rows show the median tracer--matter cross-power spectra $P_{X \times \rm m}(k)$ at $z = 0$, $1$, $2$, $3$, and $5$; the second and fourth rows show the corresponding large-scale bias $b_X(k) = P_{X \times \rm m}(k)/P_{\rm m}(k)$. Shaded bands show the 10--90 percentile range across CROCODILE realizations. Because cross-spectra contain no self-pairs, they are free of the Poisson shot noise that would otherwise dominate the BH auto-spectrum at high $k$. The stellar bias consistently exceeds the BH bias at fixed redshift in both sets. Dashed curves show the same statistic computed from the matched-volume IllustrisTNG L50 simulations, as a cross-code comparison: the CV row is compared against the IllustrisTNG CV set, and the SB26 row against a 201-run subset of the IllustrisTNG SB35 set. IllustrisTNG medians are shown without their percentile bands to limit clutter; the corresponding scatter is compared quantitatively in the text.}
    \label{fig:xpk}
\end{figure*}

A distinctive measure of how feedback shapes the large-scale clustering of different baryonic components is their cross-correlation with the total matter field. Figure~\ref{fig:xpk} shows the redshift evolution of the cross-power spectra and implied large-scale bias of stellar and black-hole mass with respect to the total matter field, for the CROCODILE L50 CV set (27 realizations) and the L50 SB26 set (201 realizations). Because BH auto-spectra are shot-noise dominated, the cross-power spectra provide a shot-noise-free measurement of BH clustering. 

A notable finding is that the large-scale stellar bias systematically exceeds the BH bias at $z = 2$: this ordering holds in all 27 out of 27 CV realizations, with a median gap of $\Delta b \approx 0.5$--$0.7$. The complete consistency across independent realizations rules out a statistical fluctuation from small BH particle numbers; were it noise, the sign would flip in at least some of the 27 phases. The effect is therefore a systematic feature of the model rather than a sampling artifact. This ordering is also recovered in the SB26 set, where it persists despite the simultaneous variation of all 26 cosmological and astrophysical parameters, indicating that it is not an artifact of the fiducial feedback tuning or cosmology but a robust feature of the model across the sampled parameter space. 

We consider three candidate physical explanations. First, at $z = 2$ the stellar mass budget may be dominated by a relatively narrow population of protocluster/group-scale halos, which are more strongly clustered than the broader halo distribution in which BH seeds are placed once the minimum FoF-mass threshold is exceeded. Second, Bondi-type accretion is numerically sensitive to the resolved gas density near the BH, so BH growth may be less tightly coupled to the large-scale environment than star formation is, effectively spreading BH mass over a wider, less biased halo population. Third, AGN self-regulation can selectively suppress further BH growth in some halos without an equivalent suppression of star formation, weakening the BH-mass to halo-mass correlation. The redshift evolution of $b_{\rm star}(z)$ and $b_{\rm BH}(z)$ visible in the figure can in principle discriminate among these scenarios.

As a first cross-code test of these trends, Figure~\ref{fig:xpk} also overlays the same statistic (dashed) computed from the 27-realization IllustrisTNG L50 CV set, matched in box size and volume to our own CV set. The stellar bias agrees closely between the two codes at every redshift shown (e.g., $b_{\rm star}(k{=}1\,h\,\mathrm{Mpc}^{-1}) = 2.4$ versus $2.0$ at $z=0$, and $3.6$ versus $3.2$ at $z=1$, for CROCODILE and IllustrisTNG respectively). The BH bias, however, differs substantially at low redshift: IllustrisTNG's BH bias is systematically lower than CROCODILE's, by $b_{\rm BH}(k{=}1\,h\,\mathrm{Mpc}^{-1}) = 1.7$ versus $2.9$ at $z=0$ and $2.1$ versus $3.6$ at $z=1$ (${\sim}40\%$ lower in both cases), with the two codes converging by $z=5$ ($6.4$ versus $6.6$). Because the stellar bias tracks so closely between the codes, this is unlikely to reflect a difference in the underlying dark-matter or halo clustering; it more plausibly reflects a difference in how the two AGN feedback and BH-seeding prescriptions populate halos across the mass function and how BHs grow at late times. 
The qualitative ordering identified above, $b_{\rm star} > b_{\rm BH}$, holds in IllustrisTNG as well, suggesting it may be a comparatively model-independent consequence of BH seeding and growth being less tightly coupled to the large-scale environment than star formation, rather than a CROCODILE-specific artifact.

The bottom rows of Figure~\ref{fig:xpk} extend this cross-code comparison to the Sobol sets, contrasting our SB26 runs against a 201-run subset of the IllustrisTNG SB35 set. As a finite-sample check, ten fixed-seed bootstrap resamples of 201 draws from this TNG subset changed the median bias by less than 10\% at $k \simeq 1~h\,\mathrm{Mpc}^{-1}$ across the redshifts and tracers shown. 
Two features are noteworthy. 
First, the low-redshift disagreement in the median BH bias largely disappears once each suite is marginalized over its own parameter prior: at $k = 1~h\,\mathrm{Mpc}^{-1}$ and $z = 0$ the Sobol medians are $b_{\rm BH} = 2.3$ (SB26) versus $1.7$ (SB35) and $b_{\rm star} = 2.4$ versus $2.3$, and by $z \gtrsim 2$ the two suites' median bias curves for both tracers track each other to within the width of the lines. The large CV-set offset therefore reflects the specific fiducial calibrations of the two models more than a generic difference between them. 
Second, the parameter-driven scatter differs systematically for the BH tracer: the 90th-to-10th-percentile ratio of $b_{\rm BH}(k{=}1\,h\,\mathrm{Mpc}^{-1})$ is $2.66$, $2.88$, and $3.99$ in SB26 at $z = 0$, $1$, and $5$, compared with $1.67$, $1.97$, and $3.35$ in SB35: the IllustrisTNG BH bias is about $1.6$ and $1.5$ times less sensitive to its parameter variations at $z = 0$ and $1$, respectively, with the gap closing toward high redshift.

\clearpage
\begin{figure*}[!t]
    \centering
    \includegraphics[width=0.95\textwidth]{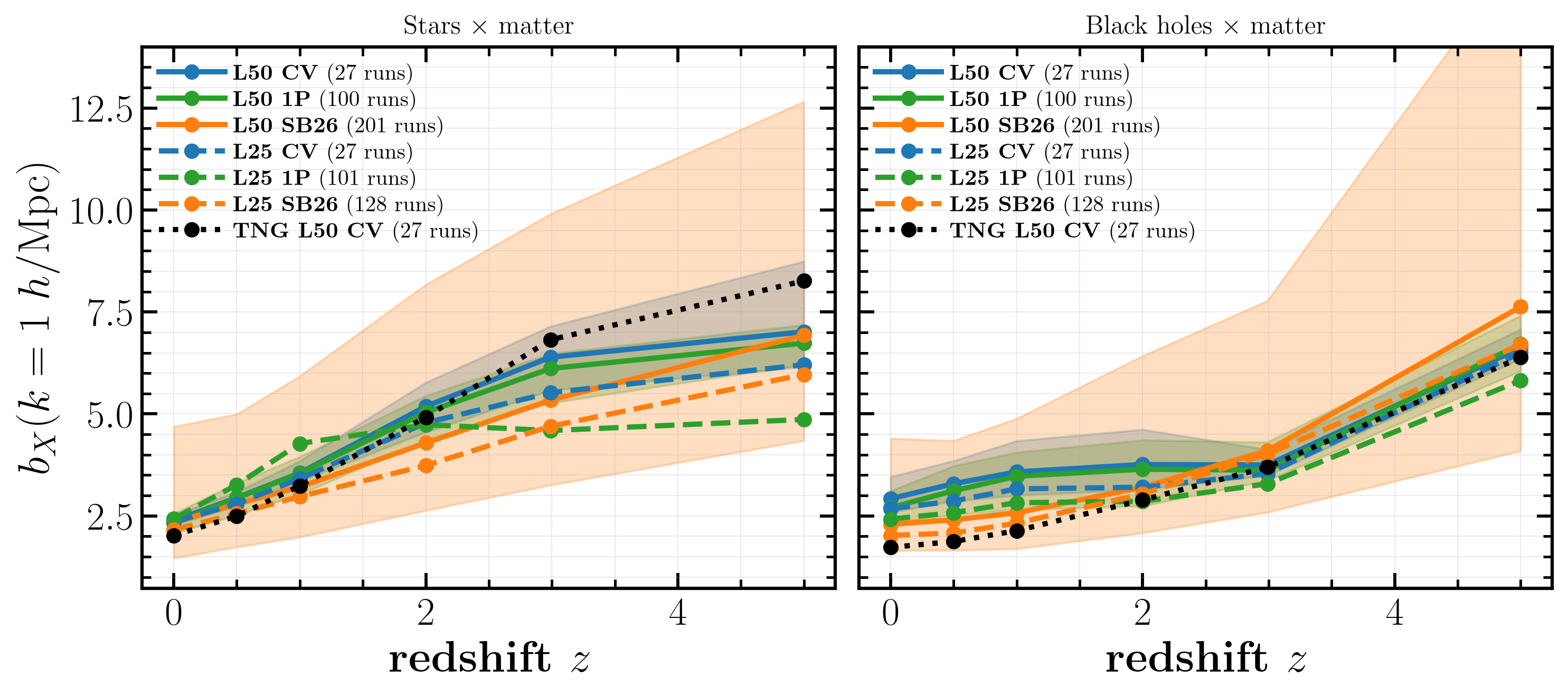}
    \caption{Large-scale bias $b_X(k = 1~h\,\mathrm{Mpc}^{-1})$ of stars (left) and black holes (right) with respect to the total matter field, as a function of redshift, for the CV (blue), 1P (green), and SB26 (orange) sets in both the L50 (solid) and L25 (dashed) boxes. Solid curves show the median and shaded bands the 10--90 percentile range across realizations for the L50 sets; the L25 medians are overplotted without shading for visual clarity, and their scatter is instead compared to the L50 scatter quantitatively in the text. The black dotted curve shows the matched-volume IllustrisTNG L50 CV set at this fixed scale, the same cross-code comparison shown across the full $k$-range in Figure~\ref{fig:xpk}.}
    \label{fig:bias_vs_z}
\end{figure*}

Figure~\ref{fig:bias_vs_z} isolates this bias evolution at a single characteristic scale, $k = 1~h\,\mathrm{Mpc}^{-1}$, showing the median and 10--90 percentile scatter as a function of redshift for the L50 CV, 1P, and SB26 sets. The medians of all three sets agree closely at every redshift for both tracers, indicating that the characteristic bias evolution is a robust feature of the fiducial model rather than an artifact of any particular parameter choice or realization. The scatter, however, differs sharply between the sets: the CV and 1P bands are comparably narrow at all redshifts, while the SB26 band widens substantially with increasing redshift, most dramatically for the BH bias, whose 10--90 percentile range grows from a factor of ${\sim}2$ at $z = 0$ to nearly a factor of $4$ by $z = 5$. Because CV isolates sample variance alone and 1P varies only one parameter at a time around the fiducial point, the near-equality of their scatter indicates that single-parameter variations do not, on their own, broaden the bias distribution beyond the level set by cosmic variance; it is only the simultaneous, joint variation of all 26 parameters in SB26 that produces substantial additional scatter, and this joint effect grows more pronounced toward high redshift, where the halo population sourcing the tracer bias is more sparsely sampled and feedback more strongly regulates the small number of star- and BH-hosting halos in each realization.

Figure~\ref{fig:bias_vs_z} also overlays the corresponding L25 medians (dashed lines) for each set. The L25 and L50 medians agree closely at low redshift for all three sets, though the L25 1P and SB26 stellar-bias medians fall increasingly below their L50 counterparts toward high redshift (e.g., $b_{\rm star} \approx 4.9$ for L25 1P versus $6.7$ for L50 1P at $z = 5$), likely reflecting the smaller box's poorer sampling of the massive, strongly star-forming halos that dominate the high-redshift stellar budget. The 10--90 percentile scatter, measured but not plotted for L25, responds oppositely in the CV and SB26 sets. In the CV set, which isolates sample variance alone, the L25 scatter is systematically wider than the L50 scatter at fixed redshift and tracer, as expected from the smaller box's larger cosmic variance: the stellar-bias 90th-to-10th-percentile ratio is $1.18$ (L25) versus $1.11$ (L50) at $z = 0$, widening to $1.70$ versus $1.42$ at $z = 5$, and the BH-bias ratio is $1.73$ versus $1.34$ at $z = 0$. This mirrors, both qualitatively and in rough magnitude, the CV-scatter reduction between the L25 and L50 boxes already noted for the SFRD in Section~\ref{sec:sfrd}. In the SB26 set, by contrast, the L25 and L50 scatter are comparable, with the L25 ratio if anything marginally smaller (stellar bias: $2.7$ versus $3.2$ at $z = 0$; BH bias: $2.1$ versus $2.7$ at $z = 0$), indicating that once all 26 parameters are varied jointly, the parameter-driven scatter dominates over the box-size-dependent cosmic-variance contribution at both box sizes, so the additional cosmic variance of the smaller box does not measurably widen the total scatter further. A small subset of the L25 SB26 realizations ($5$ of $128$) have not yet formed a measurable black-hole population by $z = 5$ under some of the sampled feedback parameters; these are excluded from the $z = 5$ BH statistic in that set.

The black dotted curve shows the IllustrisTNG L50 CV set at this same fixed scale, isolating the cross-code comparison. The stellar-bias curves track each other in shape but cross: CROCODILE lies about $18\%$ above IllustrisTNG at $z \lesssim 0.5$ ($b_{\rm star} = 2.38$ versus $2.03$ at $z = 0$), the two converge to within $5\%$ by $z = 2$, and CROCODILE falls about $15\%$ below by $z = 5$ ($7.01$ versus $8.26$). Both offsets exceed the realization-to-realization scatter of the median over $27$ realizations, so the trend is systematic rather than sample variance. The BH-bias curves diverge more strongly below $z \approx 2$, with IllustrisTNG systematically below CROCODILE before the two converge by $z = 5$.

The realization-to-realization scatter (not plotted for IllustrisTNG, for the visual-clarity) also differs between the two codes. The IllustrisTNG stellar-bias 90th-to-10th-percentile ratio is comparable to or narrower than CROCODILE's at $z \leq 2$ (e.g. $1.06$ versus $1.11$ at $z = 0$) but wider at $z = 5$ ($1.48$ versus $1.42$), while the IllustrisTNG BH-bias scatter is narrower than CROCODILE's at every redshift shown, most notably at $z = 1$--$2$ ($1.12$ versus $1.44$--$1.47$, roughly $20$--$25\%$ narrower), although the difference has largely closed by $z = 5$ ($1.08$ versus $1.11$). 
Both codes implement Bondi-like BH accretion, so this difference is not simply the presence or absence of that prescription. The seeding thresholds provide a possible explanation: every CROCODILE CV realization uses the same $M_{\rm seed}=10^5~\Msun/h$ and $M_{\rm FoF,seed}=10^{10}~\Msun/h$, whereas IllustrisTNG uses $M_{\rm seed}=8\times10^5~\Msun/h$ and $M_{\rm FoF,seed}=5\times10^{10}~\Msun/h$. CROCODILE therefore seeds smaller black holes in lower-mass halos and allows them to grow from an earlier stage; sensitivity of that initial growth to the local gas supply and assembly history could broaden the phase-to-phase BH-bias scatter. Differences in how the two AGN implementations subsequently self-regulate BH growth may contribute as well. Distinguishing these effects requires a controlled seeding comparison and is beyond the scope of the present CV-only analysis.

\subsection{Stellar-to-halo mass relation}
\label{sec:shmr}

\begin{figure*}
    \centering

    \includegraphics[width=\textwidth]{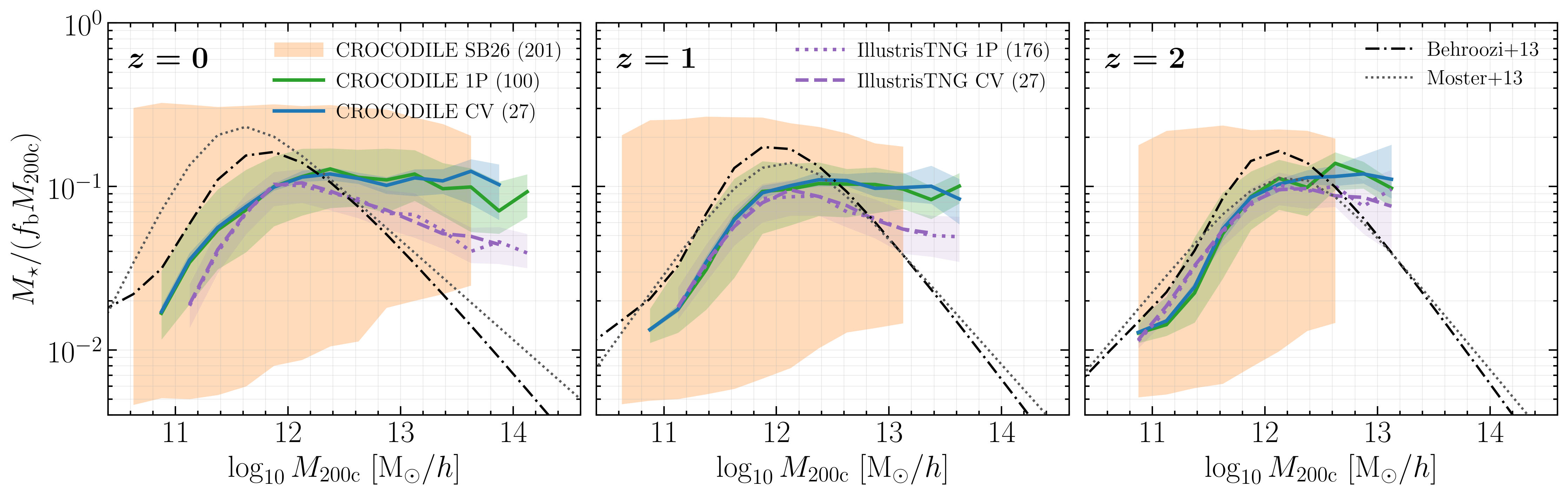}
    \caption{Stellar-to-halo mass relation of central galaxies at $z = 0$, $1$, and $2$ (left to right), for the L50 CV (blue), 1P (green), and SB26 (orange) sets, compared with the matched-volume IllustrisTNG L50 CV (purple dashed) and 1P (purple dotted) sets. Plotted is the stellar mass of the central subhalo of each FOF group normalized to the baryon budget of its host, $M_\star/(f_{\rm b} M_{\rm 200c})$.
   Each run is normalized by its own $f_{\rm b} = \Omega_{\rm b}/\Omega_{\rm m}$, which matters for the Sobol set where both parameters are varied. Thick curves show the ensemble median and shaded bands the 10--90 percentile range across runs, with the number of contributing runs given in the legend. The SB26 set is drawn as a band only: its per-bin marginal median is not the relation of any individual run and would invite over-interpretation (see text). 
  Black dash-dotted and gray dotted curves are the abundance-matching relations of \citet{Behroozi13} and \citet{Moster13}; both are calibrated against virial mass rather than $M_{\rm 200c}$, a $\lesssim 0.1$~dex horizontal shift that does not affect any statement made in the text,  and we convert their mass to $M_\odot/h$ using the fiducial $h = 0.671$. 
  Stellar mass is the total bound mass of star particles in the central subhalo, the only aperture definition that is identical in the two codes, and consequently includes diffuse intrahalo light (see text). Bins are shown only where the median central stellar mass exceeds $2\times10^{8}~\Msun$ (${\approx}21$ star particles at the fiducial spawn number) in at least $90\%$ of the runs of a given set.   }
    \label{fig:shmr}
\end{figure*}

The cosmic star formation history of Section~\ref{sec:sfrd} measures the volume-integrated efficiency of star formation, but says nothing about which halos the stars form in. The stellar-to-halo mass relation (SHMR) resolves that integral in halo mass and is the most direct summary of how a feedback model distributes star formation across the halo population; it is also the quantity against which galaxy formation models are commonly calibrated. Figure~\ref{fig:shmr} shows it for the central galaxies of the L50 suite at $z = 0$, $1$, and $2$, alongside the matched-volume IllustrisTNG L50 sets. For every FOF group we take the central subhalo identified by \textsc{Subfind} and pair its stellar mass against the group's $M_{\rm 200c}$. Stellar mass used here is the \emph{total bound} mass of star particles. 
Masses are converted from the $10^{10}\,h^{-1}\Msun$ convention using each run's own $h$, which is itself a varied parameter.

Two features stand out. First, the two codes agree closely where stellar feedback sets the efficiency. For $M_{\rm 200c} \lesssim 10^{12.2}~\Msun/h$, which brackets the peak of the abundance-matching relations, the CV medians of the two suites agree to within $10\%$ at every redshift shown.
Given that the two suites share nothing but their cosmology, volume, and mass resolution, and reach this efficiency through structurally different stellar feedback (locally calibrated mechanical injection against a star-formation-driven kinetic wind whose launch velocity is tied to the local velocity dispersion and whose energy has a metallicity-dependent scaling), this level of agreement is a non-trivial validation of the Osaka model at the mass scale where most of the cosmic stellar mass resides.

Second, the two codes diverge systematically toward higher halo mass, in the regime where AGN feedback rather than stellar feedback regulates growth. The ratio of the CROCODILE to the IllustrisTNG median efficiency grows from $1.38$ at $10^{12.6}~\Msun/h$ to $1.86$ at $10^{13.1}~\Msun/h$ and $2.50$ at $10^{13.6}~\Msun/h$ at $z=0$, with closely similar behavior at $z=1$.
Equivalently, the IllustrisTNG efficiency turns over at $M_{\rm 200c} \simeq 10^{12.1}~\Msun/h$ and declines steeply thereafter, whereas the CROCODILE relation rises to a broad plateau above $10^{12.3}~\Msun/h$ and stays flat out to the largest halos the box contains. The physical reading is that CROCODILE's purely thermal AGN energy injection does not suppress star formation in group-scale central galaxies as efficiently as IllustrisTNG's kinetic radio-mode feedback. This is the same model difference that Section~\ref{sec:pk_suppression} exposes from a completely different direction: the CROCODILE fiducial model suppresses the matter power spectrum by only a few percent where IllustrisTNG suppresses it by ${\sim}27\%$ at $z=0$. Both measurements indicate that CROCODILE's AGN feedback removes and redistributes far less baryonic material from massive halos, and it is notable for the interpretation of both that the halo mass at which the SHMR curves separate, $M_{\rm 200c} \gtrsim 10^{12.3}~\Msun/h$, is also the mass scale whose baryon content controls $S(k)$ in the \citet{vanDaalen20} picture.

An important caveat attaches to the massive end. Because the comparison uses total bound stellar mass, it includes the diffuse intrahalo light, which is not counted in the aperture-limited stellar masses that the \citet{Behroozi13} and \citet{Moster13} relations were calibrated against. Both suites therefore lie above those relations for $M_{\rm 200c} \gtrsim 10^{12.3}~\Msun/h$, and the size of that offset should not be read as a pure model failure. The \emph{difference between the two codes} is immune to this, since both are measured with the identical definition.

The scatter across the three CROCODILE sets reproduces the hierarchy seen in the other statistics presented here. At $M_{\rm 200c} \approx 10^{12.1}~\Msun/h$ and $z=0$, the 90th-to-10th-percentile range of $M_\star/(f_{\rm b}M_{\rm 200c})$ is a factor of $1.17$ across the CV set, $2.6$ across the 1P set, and $23$ across the SB26 set. The CV band, which isolates sample variance at fixed physics, is far narrower than the IllustrisTNG-to-CROCODILE offset at high mass, confirming that the cross-code divergence is a model difference and not a realization fluctuation. 
The SB26 band is very wide,
and it brackets both fiducial models at every mass and redshift shown, and no median curve is drawn for the SB26 set. 
Plotting that curve would suggest a characteristic Sobol relation that no simulation in the set actually follows, so we show the percentile band alone, which is in any case the quantity of interest for the prior. That is precisely the property the suite is built to have: an ML model trained on the CAMELS-CROCODILE Sobol set sees a range of star formation efficiencies wide enough to contain the IllustrisTNG behavior as an interior point rather than an extrapolation, which is a prerequisite for the cross-suite robustness tests discussed in Section~\ref{sec:outlook}.

\subsection{Lyman-$\alpha$ absorption around galaxies}
\label{sec:lya_decrement}

\begin{figure*}
    \centering
    \includegraphics[width=\textwidth]{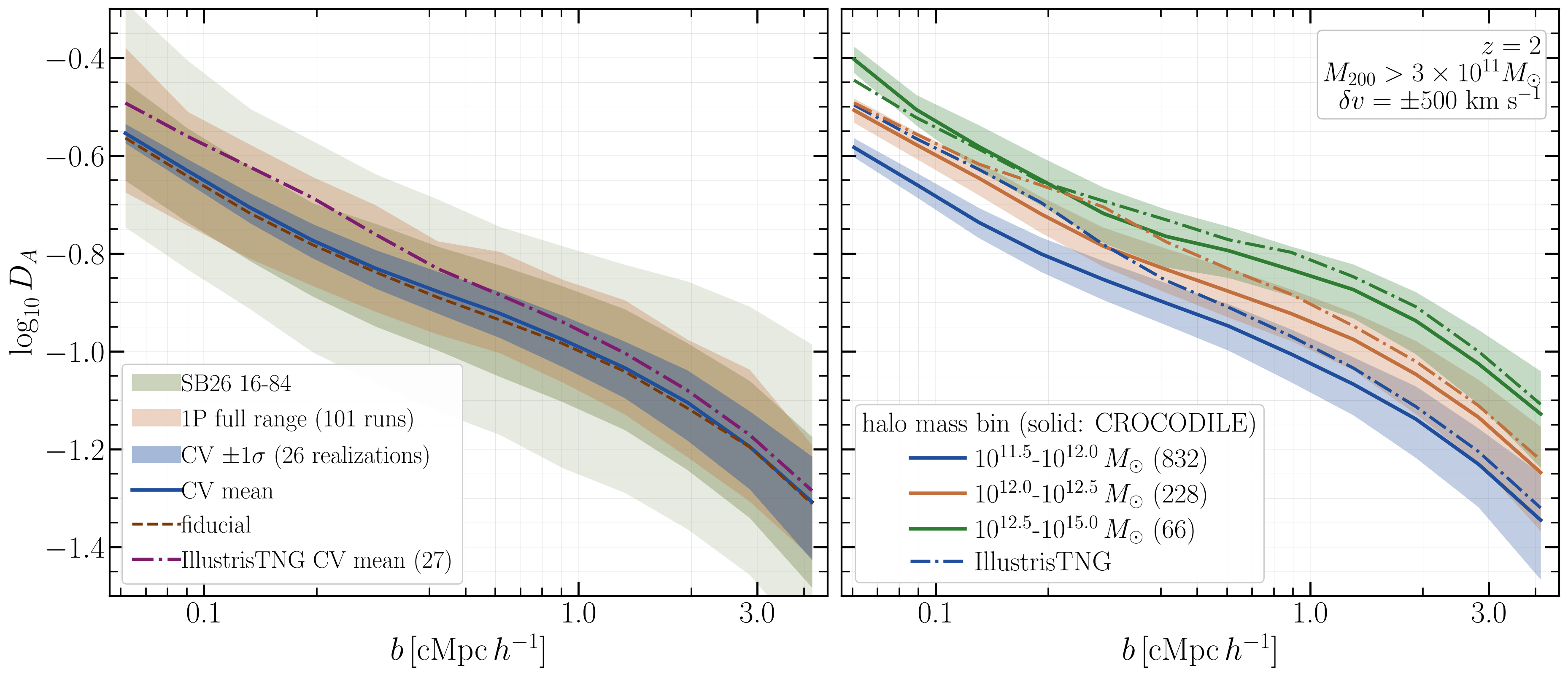}
    \caption{Mean Ly$\alpha$ flux decrement $D_A$ around the central galaxies of $M_{200} > 3\times10^{11}\,M_\odot$ halos at $z = 2$ in the L50 suite, using a $\pm500$ km s$^{-1}$ line-of-sight velocity window. Left: all selected halos pooled. The blue band is the $\pm1\sigma$ scatter over the $26$ CV realizations, which is the sample variance due to seed difference (i.e. cosmic variance). 
    The solid blue line is the CV mean and the dashed line the fiducial 1P run. The orange band spans the full range of the $101$ 1P runs, in which one parameter at a time is varied, and the green bands show the $16$--$84$ and $2$--$98$ percentile ranges of the $201$ completed SB26 runs, which sample the full $26$-dimensional parameter space. The parameter-driven spread exceeds the cosmic variance by roughly a factor of eight at $b \approx 60\,h^{-1}$ ckpc but falls to unity by $b \approx 4\,h^{-1}$ cMpc, so the feedback information is concentrated at $b \lesssim 1\,h^{-1}$ cMpc. 
    Right: the identical measurement split into three halo-mass bins, showing the CV mean and its $\pm1\sigma$ realization scatter, with the median number of halos per realization in parentheses. The decrement is ordered by halo mass at every impact parameter, which motivates the scaling applied in Figure~\ref{fig:lya_DA_r200}. The dot-dashed line in both panels is the mean over the $27$ matched-volume IllustrisTNG L50 CV realizations, measured with the same selection, sightlines, velocity window and mean-flux normalization; its realization scatter is comparable to CROCODILE's and is omitted to limit clutter.}
    \label{fig:lya_DA}
\end{figure*}

A distinctive diagnostic of how feedback shapes the circumgalactic and intergalactic gas is the Lyman-$\alpha$ (Ly$\alpha$) flux decrement in background-quasar or background-galaxy sightlines as a function of the transverse impact parameter $b$ from foreground galaxies. Observationally, the mean Ly$\alpha$ absorption around $z \sim 2$--$3$ star-forming galaxies has been measured from stacked sightline samples \citep{Steidel10,Rakic12,Turner14,Chen20} and from IGM tomography and galaxy--forest cross-correlations \citep{Mukae17,Momose21a,Momose21c}, and \citet{nagamine2021feedback} showed with \texttt{GADGET3-Osaka} that the decrement profile at $b \lesssim 1$ comoving Mpc is directly sensitive to the treatment of stellar and AGN feedback, as well as the UV background radiation field.

A related analysis in a different subgrid framework reached a qualitatively similar conclusion about feedback sensitivity: \citet{sorini2020simba} extracted the mean Ly$\alpha$ absorption profile over $10~{\rm kpc} \lesssim b \lesssim 10~{\rm Mpc}$ around $2 \leq z \leq 3$ quasars in SIMBA and, by comparing runs with individual feedback modules deactivated, identified stellar feedback rather than mechanical AGN feedback as the primary driver of the average CGM properties on these scales. That measurement is not directly comparable to the one we describe below, since it is centred on quasars rather than on a stellar-mass-selected galaxy sample and uses a different hydrodynamic scheme, feedback implementation, and box configuration, so we draw no quantitative comparison with it. Its relevance here is instead that it independently demonstrates the diagnostic power of the statistic.  \citet{sorini2020simba} also found that SIMBA predicts appreciably more absorption within $100$ kpc (despite its strong jet feedback) than the \textsc{nyx} and \textsc{illustris} calculations they compared against. Whatever the origin of that particular offset, the fact that independent models disagree at this level indicates that the absorption profile discriminates between feedback prescriptions rather than being a statistic on which all models converge. Neither the second-generation IllustrisTNG suite \citep{genel2026camels50} nor the ASTRID extension \citep{ni2023camels} examined this galaxy--absorption cross-correlation, making it a natural axis along which CAMELS-CROCODILE adds new information; within CAMELS, IGM studies have so far focused on the line-of-sight forest statistics alone \citep{Tillman23a,Tillman25}.

Here we measure the mean Ly$\alpha$ flux decrement profile $D_A(b) = 1 - \langle F(b) \rangle / \langle F_{\rm IGM} \rangle$ around the central galaxies of halos above a fixed halo mass threshold, following the methodology of \citet{nagamine2021feedback}, and exploit the CAMELS design in two ways that go beyond that work. 
The 1P set provides controlled variations with which the response of $D_A(b)$ to individual feedback parameters can be studied in future work, while the SB set supplies a broad training set for applying this statistic to emulator or machine-learning inference. In this sense, the galaxy--forest cross-correlation offers a complementary observable to the matter power spectrum, with sensitivity concentrated on circumgalactic scales.
The larger L50 volume is important here, as it contains the more massive halos around which the observed samples are centered and reduces the sample variance of the cross-correlation on transverse scales of several Mpc.

Figure~\ref{fig:lya_DA} shows the resulting profiles at $z = 2$. Ly$\alpha$ optical depths are computed with \texttt{fake\_spectra} \citep{Bird17fake} from the simulation's own non-equilibrium \texttt{Grackle} neutral-hydrogen field rather than from a photoionization-equilibrium fitting formula, the same choice made by \citet{Tillman23} for the first-generation CAMELS suites, and are rescaled in each realization separately so that the box mean flux matches the observed value at this redshift \citep{Faucher08a}. 
Sightlines are placed directly around the selected galaxies, eight per halo in each of twelve logarithmic impact-parameter bins between $50$ and $5000\,h^{-1}$ ckpc, with the radius drawn uniformly in annulus area so that each bin average is unbiased. 
The profiles begin at the center of the innermost impact-parameter bin, $b=61.7\,h^{-1}$\,ckpc, with sightlines sampled in logarithmic annuli spanning $50$--$5000\,h^{-1}$ ckpc.
Here we do not overlay the observational data points deliberately, because the purpose of this paper is to document the suite as a simulation dataset, including for ML inference, rather than to make a precision comparison against data, which would require careful selection effect considerations.

We select on halo mass rather than on stellar mass because a fixed $M_\star$ cut is not a fixed population across the suite: the number of galaxies above $10^{10}\,M_\odot$ varies by a factor of $0.06$ to $3.8$ between 1P runs as feedback moves the stellar masses, and the apparent $D_A$ response then correlates with that change
rather than tracking the gas alone. Selecting central galaxies of halos above $3 \times 10^{11}\,M_\odot$ holds the population to within $0.83$--$1.13$ of the fiducial count over the same runs. The velocity window is set by the redshift precision expected for the foreground sample: \citet{Greene22} find that $90\%$ of simulated PFS galaxies recover $\Delta z/(1+z) < 5\times10^{-4}$, or $150$ km s$^{-1}$, so a $\pm500$ km s$^{-1}$ window comfortably contains the sample while retaining substantially more signal than the $\pm1000$ km s$^{-1}$ adopted by \citet{nagamine2021feedback}, who assumed a larger $500$ km s$^{-1}$ redshift scatter.

The decrement falls from $D_A = 0.279 \pm 0.013$ at $b \approx 62\,h^{-1}$ ckpc to $0.049 \pm 0.012$ at $b \approx 4.2\,h^{-1}$ cMpc, where the quoted uncertainty is the CV realization scatter. Two features of Figure~\ref{fig:lya_DA} are relevant for inference. First, the parameter-driven spread exceeds the sample variance by a factor of $8.1$ at the innermost bin but only $1.1$ at the outermost, so a single L50 volume cannot distinguish parameter variations from cosmic variance beyond a few comoving Mpc regardless of measurement precision; the constraining power of this statistic resides at $b \lesssim 1\,h^{-1}$ cMpc. Second, the SB26 median tracks the CV mean to within $0.8\sigma$ of the sample variance at every impact parameter, confirming that the Sobol sample is centered on the fiducial model rather than on some displaced region of parameter space, which is a prerequisite for using it as emulator training data.

Figure~\ref{fig:lya_DA} also shows the same measurement applied to the IllustrisTNG L50 CV set, as a cross-code comparison. Everything except the galaxy-formation physics is held fixed: the same box size and cosmology, the same $z \simeq 2$ output,
the same selection of central galaxies of $M_{200} > 3\times10^{11}\,M_\odot$ halos ($1234$ of them in the fiducial TNG realization against $1174$ in CROCODILE), the same sightline geometry and velocity window, and the same per-realization rescaling to the observed mean flux. As for CROCODILE, the neutral hydrogen is taken from the simulation itself, which for TNG means the \texttt{NeutralHydrogenAbundance} field; we verified on one realization that whether or not the star-forming gas is additionally forced neutral changes $D_A$ by at most $0.5\%$ at the impact parameters considered here, since the star-forming gas lies well inside the innermost bin. IllustrisTNG produces systematically more absorption than CROCODILE, by $15$--$22\%$ at $b < 200\,h^{-1}$ ckpc, with the offset peaking near the virial radius and decaying to $5\%$ and statistical insignificance beyond $b \sim 1\,h^{-1}$ cMpc. Measured against the CROCODILE scatter, the difference reaches $3.4\sigma$ in the innermost bin. Because the two suites use independent initial conditions, roughly $1\sigma$ of that offset is simply the difference between two random volumes, so we regard the disagreement at $b \lesssim 200\,h^{-1}$ ckpc as significant and the smaller offsets at larger $b$ as not. We also note that the two codes necessarily differ in how the neutral fraction is computed, non-equilibrium \texttt{Grackle} chemistry in CROCODILE against photoionization equilibrium in TNG, and this difference is part of what the comparison measures and cannot be separated from the feedback difference. With that caveat, the result is a direct demonstration of the point made above using \citet{sorini2020simba}: the absorption profile around galaxies is a statistic on which independent feedback models disagree at the tens of per cent level on circumgalactic scales, rather than one on which they converge.

Among the explored 1P variations, the response is dominated by parameters controlling the star-formation threshold and the energetics of stellar-driven winds, and is confined to small impact parameters. Ranked by peak significance relative to the within-realization jackknife error, the strongest responses are the star-formation density threshold ($23\sigma$, a $53\%$ change in $D_A$), the SNII energy boost factor ($12\sigma$, $23\%$), the wind velocity boost ($8\sigma$, $15\%$) and the wind sound-speed boost ($6\sigma$, $10\%$); all four peak in the innermost bin at $b \approx 62\,h^{-1}$ ckpc and decay by $b \sim 200\,h^{-1}$ ckpc. Lowering the star-formation density threshold raises the decrement, while boosting the wind energetics lowers it, as expected if stronger winds evacuate neutral gas from the inner circumgalactic medium. The wind mass loading, by contrast, produces no significant response once the galaxy population is held fixed ($1.8\sigma$), so it suggests that the energy and velocity of the outflows (rather than their mass budget) shape the profile. All six parameters governing black hole accretion and AGN feedback give peak responses below $1.4\sigma$, and the scatter they induce is a factor of $3$ to $11$ smaller than the CV sample variance, so AGN feedback leaves no detectable imprint on $D_A(b)$ at $z = 2$ in this model. This mirrors the conclusion \citet{sorini2020simba} reached for SIMBA by a different route, and it should be read as a statement about $z = 2$ halos of this mass in a $50\,h^{-1}$ cMpc volume rather than as a general claim about AGN and the CGM.

\begin{figure}[!t]
    \centering
    \includegraphics[width=\columnwidth]{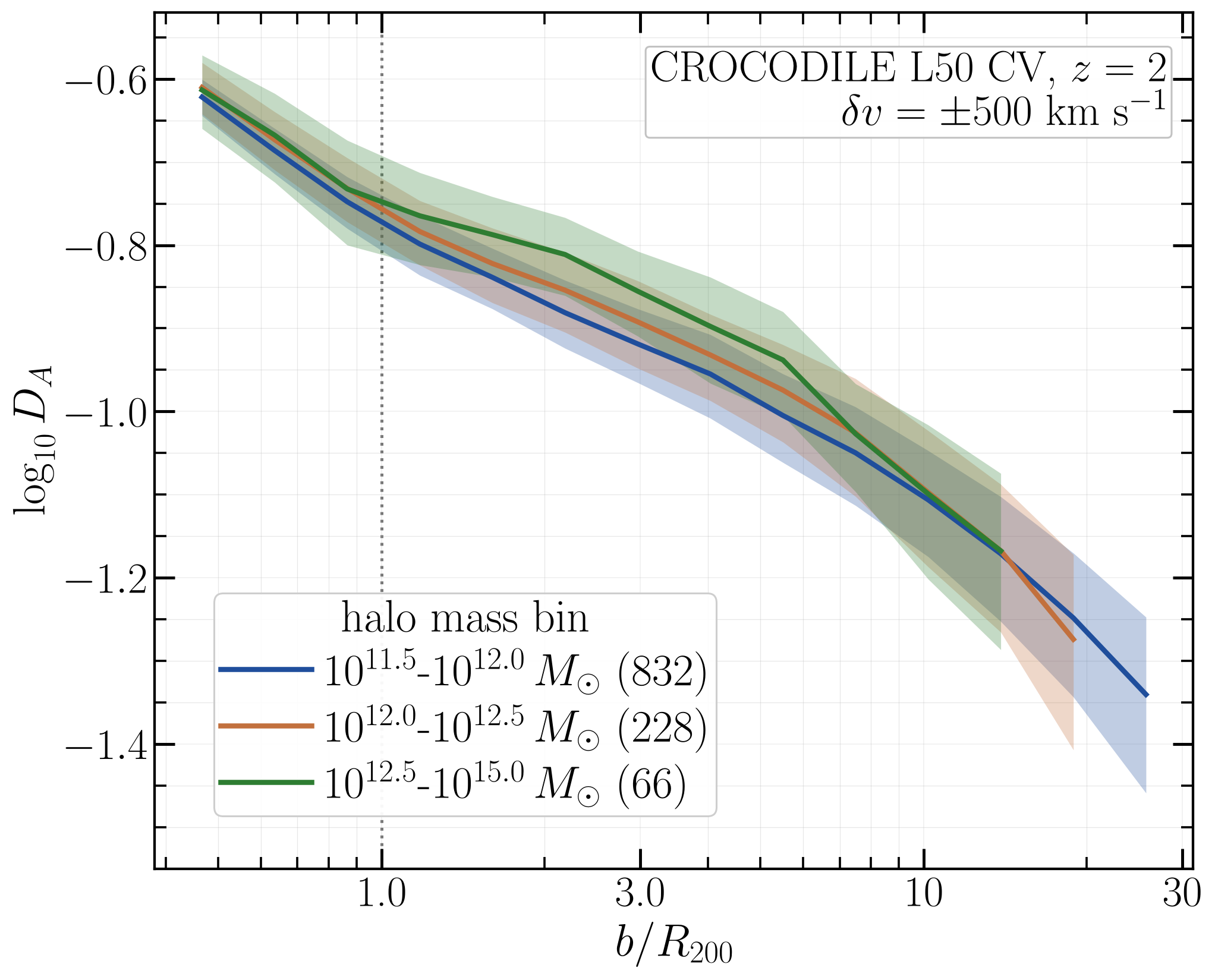}
    \caption{The same halo-mass-split measurement as the right panel of Figure~\ref{fig:lya_DA}, but with the impact parameter of every sightline divided by the virial radius $R_{200}$ of its own host halo before stacking. Lines and shading are the CV mean and its $\pm1\sigma$ realization scatter; the dotted line marks $b = R_{200}$. The mass ordering seen in Figure~\ref{fig:lya_DA} largely disappears.
    Each curve is truncated where its own mass bin runs out of sightlines: because the sightlines stop at a fixed $5\,h^{-1}$ cMpc, the most massive halos, which have the largest $R_{200}$, cannot reach large $b/R_{200}$.}
    \label{fig:lya_DA_r200}
\end{figure}

The right panel of Figure~\ref{fig:lya_DA} shows that the decrement is ordered by halo mass at every impact parameter, the more massive halos producing more absorption simply because they are physically larger and have more neutral hydrogen. 
Figure~\ref{fig:lya_DA_r200} repeats that measurement with the impact parameter of each sightline divided by $R_{200}$ of its own host halo, normalising per halo rather than per mass bin. The mass dependence largely disappears: over a matched dynamic range the mean profiles of the three bins differ by $0.157$ dex at fixed $b$ but by only $0.047$ dex at fixed $b/R_{200}$, a reduction of a factor of $3.3$, and near the virial radius the three bins agree to within the realization scatter. \citet{Meiksin17} likewise found that scaling the impact parameter by the virial radius yields nearly universal H~I absorption profiles across halo mass in Sherwood, both with and without winds; the wind-induced enhancement is also nearly mass independent in those units. This provides an independent precedent for the collapse in Figure~\ref{fig:lya_DA_r200}, while our 1P suite identifies the feedback controls that set its amplitude. The absorption around these galaxies is therefore close to self-similar in units of the virial radius, so it is the profile shape rather than its amplitude that is the mass-independent quantity. 
The scaled version isolates the response of the CGM to varying feedback models, whereas the unscaled version is the one that could be confronted with observations, since observed samples are selected by stellar mass or luminosity and do not carry halo masses.

\section{Cross-code robustness of graph-based cosmological inference}
\label{sec:gnn}

The preceding sections characterized CAMELS-CROCODILE through conventional summary statistics. We close with a first demonstration of the use for which the suite was built: as an \emph{out-of-distribution test set} for machine-learning models trained on other galaxy formation implementations. Robustness of graph-based inference to subgrid physics has so far been tested for dark matter halos, with networks trained on $N$-body catalogs \citep{shao2023halos}, and for galaxies, with networks trained on the IllustrisTNG, SIMBA, or ASTRID suites and tested on the others and on Magneticum and SWIFT-EAGLE sets \citep{desanti2023galaxies}; for galaxies, only the ASTRID-trained network proved robust, and cross-suite tests with field maps show a similar model dependence \citep{ni2023camels}. CAMELS-CROCODILE extends such tests to an independent feedback model calibrated to local superbubble physics and to $(50~\Mpch)^3$ volumes, and because it shares the volume, mass resolution, and catalog format of the second-generation IllustrisTNG suite \citep{genel2026camels50}, a network trained there can be tested with no adaptation at all.

\subsection{Experimental design}
\label{sec:gnn_setup}

We take the graph neural network (GNN) that \citet{genel2026camels50} trained on galaxy graphs from their SB35 set to infer $\Omega_{\rm m}$ and $\sigma_8$ from the positions and line-of-sight velocities of galaxies \citep[following][]{villanueva2022graphs,desanti2023galaxies}, and apply it to CAMELS-CROCODILE with its weights \emph{frozen}: no retraining, fine-tuning, or recalibration is performed. The graphs are built with the identical prescription: galaxies are linked to all neighbors within $r_{\rm link} = 2.15~\Mpch$, edges carry the invariant features $\{|\mathbf{d}_{ij}|/r_{\rm link}, \alpha_{ij}, \beta_{ij}\}$, and each node carries its $z$-direction peculiar velocity.

The galaxy selection must match the training set, and here one point requires care. \citet{genel2026camels50} retain subhalos with at least $20$ star particles, but this criterion does not transfer to CAMELS-CROCODILE, because the number of star particles spawned per gas particle, $n_{\rm spawn}$, is one of the varied parameters (Table~\ref{tab:params}). The stellar particle mass is $m_{\rm gas}/n_{\rm spawn}$, so a fixed particle count would correspond to a stellar mass that varies by a factor of four across the Sobol set and would correlate the galaxy sample with an astrophysical parameter. We therefore select on stellar mass. The median stellar particle mass measured in the IllustrisTNG L50 CV catalogs at $z=0$ is $9.7\times10^6~\Msun/h$, so the 20-particle training selection corresponds to $M_\star > 1.94\times10^8\,(\Omega_{\rm b}/0.049)~\Msun/h$, which we adopt; this is somewhat above the value of $1.5\times10^8$ mentioned by \citet{genel2026camels50}, but we confirmed that they have adopted the $>20$ particle criteria. 
The CROCODILE runs with $n_{\rm spawn}=1$ have roughly the same stellar particle mass as IllustrisTNG. 
We use all runs ($n_{\rm spawn}=1-4$) rather than just the $n_{\rm spawn}=1$ subset. 

With this selection the fiducial CV runs contain a median of $6.1\times10^3$ galaxies per box, against $5.1\times10^3$ in the IllustrisTNG CV graphs built with the training selection, so at fixed physics the CROCODILE graphs are about $20\%$ larger than the training graphs. The size of this excess depends strongly on $n_{\rm spawn}$, which acts on the galaxy population physically and not only through resolution: in the 1P runs, which vary it alone at fixed cosmology and seed, the number of galaxies above the threshold rises from $5.1\times10^3$ to $9.2\times10^3$ between $n_{\rm spawn}=1$ and $4$, against a seed-to-seed scatter of $2\%$, and the stellar mass function changes most at $M_\star \sim 10^{9.5}~\Msun/h$, where galaxies contain several hundred to over a thousand star particles.  We observe an increase of the lowest mass galaxies in the case of $n_{\rm spawn}=4$ and the change in the galaxy stellar mass function is non-uniform. 
A plausible mechanism is the granularity of the feedback: for a fixed star formation rate, a larger $n_{\rm spawn}$ produces more numerous but lighter star particles, which inject supernova energy in smaller packets that couple less effectively to the surrounding gas, weakening the regulation of star formation. No value of $n_{\rm spawn}$ reproduces the IllustrisTNG stellar mass function perfectly, however, at $n_{\rm spawn}=1$ the count above the threshold agrees with IllustrisTNG to $2\%$ only because an excess just above the threshold offsets a deficit of $15$--$60\%$ at $10^{9}$--$10^{10.5}~\Msun/h$.

\subsection{Results}
\label{sec:gnn_results}

\begin{figure*}
    \centering
    \includegraphics[width=\textwidth]{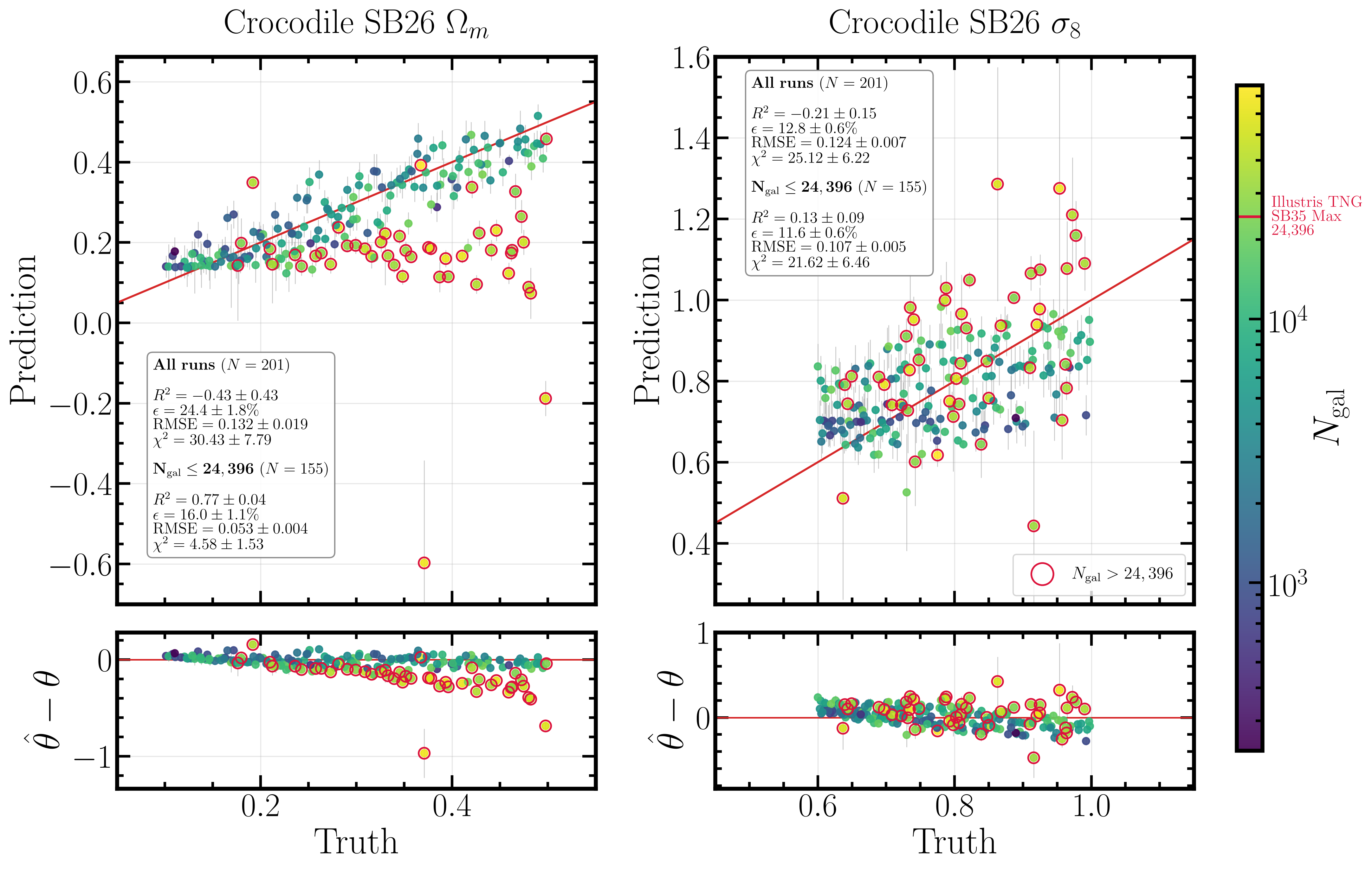}
    \caption{Inference of $\Omega_{\rm m}$ (left) and $\sigma_8$ (right) for the \NSBrunsGNN{} CAMELS-CROCODILE L50 Sobol runs by the galaxy-graph GNN of \citet{genel2026camels50}, trained on the IllustrisTNG SB35 set and applied with its weights frozen, using galaxies with $M_\star > 1.94\times10^8\,(\Omega_{\rm b}/0.049)~\Msun/h$. Each point is one simulation, colored on a logarithmic scale by the number of galaxies $N_{\rm gal}$ in its graph, with the vertical bar showing the posterior standard deviation predicted by the network; the red line marks perfect recovery. Open red circles mark the $46$ runs whose graphs contain more galaxies than any of the $1024$ IllustrisTNG SB35 graphs ($N_{\rm gal} > 24{,}396$, indicated on the color bar). The lower panels show the residual $\hat{\theta}-\theta$ against the true value. The insets give, for all runs and for the $155$ runs with $N_{\rm gal} \leq 24{,}396$, the coefficient of determination $R^2$, the mean relative error $\epsilon$, the root-mean-square error (RMSE), and the mean $\chi^2$ per simulation of the residuals in units of the predicted uncertainty; following the SB35 analysis, runs with $\chi^2 > 1000$ are excluded from the $\chi^2$ average but all runs are plotted. On its in-distribution SB35 test set the same network achieves an RMSE of $0.035$ for $\Omega_{\rm m}$ and $0.093$ for $\sigma_8$.}
    \label{fig:gnn}
\end{figure*}

Figure~\ref{fig:gnn} shows that the network does not transfer. For $\Omega_{\rm m}$ the RMSE rises from $0.035$ on the in-distribution SB35 test set to $0.132 \pm 0.019$ on CAMELS-CROCODILE, nearly a factor of four, with a mean relative error of $24\%$; for $\sigma_8$ it rises from $0.093$ to $0.124 \pm 0.007$, a factor of $1.3$. The coefficient of determination is consistent with or below zero for both parameters ($R^2 = -0.43 \pm 0.43$ and $-0.21 \pm 0.15$), so the network recovers the true values no better than a constant equal to their mean. The failure is not simply added noise: the predictions are compressed toward the center of the training range, lying above the truth at low $\Omega_{\rm m}$ and $\sigma_8$ and below it at high values, as expected of a posterior-mean estimator that finds its input uninformative and falls back toward its prior, and several runs with true $\Omega_{\rm m} \approx 0.35$--$0.5$ are catastrophic outliers, with inferred values of $0.07$--$0.12$ and in one case below zero. 
In a test on the first $128$ runs, selecting galaxies at the looser threshold of $1.5\times10^8\,(\Omega_{\rm b}/0.049)~\Msun/h$ instead changed the RMSE by at most $5\%$ and left the other metrics within their uncertainties, so these conclusions do not depend on the choice.

The failure is concentrated in the runs whose graphs lie beyond the training range in galaxy number. Restricting the test to the $155$ runs with no more galaxies than the largest SB35 graph ($N_{\rm gal} \leq 24{,}396$) reduces the $\Omega_{\rm m}$ RMSE from $0.132$ to $0.053 \pm 0.004$, $1.5$ times the in-distribution value, raises $R^2$ to $0.77 \pm 0.04$, and lowers $\chi^2$ to $4.6 \pm 1.5$; the catastrophic outliers are essentially all among the $46$ runs beyond that range (open circles in Figure~\ref{fig:gnn}). For $\sigma_8$ the same restriction helps far less: the RMSE falls only to $0.107 \pm 0.005$, $1.15$ times the in-distribution value, $R^2$ reaches only $0.13 \pm 0.09$, and $\chi^2$ remains $22 \pm 6$. The extrapolation in galaxy number therefore accounts for most of the $\Omega_{\rm m}$ failure, but not for the overconfident $\sigma_8$ posteriors.

The more consequential failure concerns the uncertainties. The posterior widths predicted by the network do not grow to reflect its errors: the mean $\chi^2$ is $30 \pm 8$ for $\Omega_{\rm m}$ and $25 \pm 6$ for $\sigma_8$, far above the value of order unity expected for calibrated uncertainties. For consistency with the SB35 analysis, these are truncated means that exclude the three runs per parameter with $\chi^2 > 1000$; all runs remain shown in Figure~\ref{fig:gnn}, and every excluded run has a galaxy number beyond the training range. The corresponding untruncated means are $56$ for $\Omega_{\rm m}$ and $265$ for $\sigma_8$, so the conclusion of severe miscalibration does not depend on this convention. The network is therefore not merely inaccurate but confidently so, returning narrow posteriors that exclude the true parameters, with nothing internal to the inference to signal that its input lies outside the training distribution. Applied to observations, where no ground truth reveals the problem, this is the more dangerous of the two failure modes.

The degradation is also strongly parameter-dependent. In distribution the network constrains $\Omega_{\rm m}$ considerably better than $\sigma_8$; on CAMELS-CROCODILE the two errors are comparable, so the degradation is far larger for $\Omega_{\rm m}$, a factor of $3.8$ against $1.3$ for $\sigma_8$. Most of the additional $\Omega_{\rm m}$ error comes from the runs beyond the training range in galaxy number, whereas $\sigma_8$ is degraded across the whole test set, a pattern that the controlled tests described below help to explain.

Splitting the test set by $n_{\rm spawn}$ tests whether the degradation is a numerical effect of graph size, which grows steadily with $n_{\rm spawn}$ (Section~\ref{sec:gnn_setup}). The runs with $n_{\rm spawn}=1$, whose stellar particle mass roughly matches IllustrisTNG, recover $\Omega_{\rm m}$ best, with an RMSE of $0.095 \pm 0.015$ and the only positive $R^2$ ($0.29 \pm 0.24$), against $0.117 \pm 0.036$, $0.164 \pm 0.053$, and $0.129 \pm 0.020$ for $n_{\rm spawn} = 2$, $3$, and $4$, but there is no trend beyond that: the $\Omega_{\rm m}$ RMSE does not change monotonically, its $\chi^2$ is essentially flat ($33$, $31$, $35$, and $21$), and the $\sigma_8$ metrics show no ordering at all. With $46$--$53$ runs per subset the differences are at most at the $1.5\sigma$ level, and the subsets of the Sobol design also differ in their feedback parameters, so the split gives no evidence that graph size really drives the failure.

Controlled tests on the CV and 1P sets clarify how galaxy number enters the inference. For the $27$ CV runs, which share the fiducial parameters and contain about $6\times10^3$ galaxies each, well inside the SB35 range, the network recovers the fiducial cosmology with small mean biases, $\Delta\Omega_{\rm m} = +0.013 \pm 0.007$ and $\Delta\sigma_8 = +0.028 \pm 0.011$, and its predicted uncertainties are well calibrated there, with mean $\chi^2$ of $1.1$ for $\Omega_{\rm m}$ and $1.4$ for $\sigma_8$. For the 1P runs that vary $\Omega_{\rm m}$ alone, the inferred values of $0.14$, $0.24$, $0.34$, $0.40$, and $0.45$ track the true values of $0.1$--$0.5$, with the compression toward the prior noted above. The 1P $n_{\rm spawn}$ sweep then isolates galaxy number itself: at fixed cosmology and seed, raising the galaxy count from $5.1\times10^3$ to $9.2\times10^3$ lowers the inferred $\Omega_{\rm m}$ from $0.34$ to $0.31$ and raises $\sigma_8$ from $0.74$ to $0.89$, a shift of about $2.5$ times the predicted $\sigma_8$ uncertainty. The network therefore does not read additional galaxies as additional matter. It reads them predominantly as a higher fluctuation amplitude, with $\Omega_{\rm m}$ moving the opposite way.

Finally, we ask where in the CAMELS-CROCODILE parameter space the failures lie. The runs beyond the training range occupy a specific corner of the Sobol design: of the $26$ parameters, only five differ significantly between them and the remainder (Kolmogorov--Smirnov tests with a false-discovery-rate correction), namely weak supernova feedback (in the Type~II SN energy boost, the SN momentum boost $A_{\rm SN1}$, and the hypernova fraction), together with high $\Omega_{\rm m}$ and $\sigma_8$. Their effect on the inference, however, is mediated by galaxy number. The $\Omega_{\rm m}$ residual correlates with $A_{\rm SN1}$ and the Type~II SN energy boost (Spearman $\rho = +0.28$ and $+0.27$), but both correlations vanish once $\log N_{\rm gal}$ is controlled for ($-0.00$ and $-0.12$), leaving only the cosmological parameters, whose residual trends are the compression toward the prior described above. Among the in-range runs, the quarter with the largest $\Omega_{\rm m}$ errors is statistically indistinguishable from the rest in every one of the $26$ parameters, and a model of the error that already uses $\log N_{\rm gal}$ is not improved by adding them. Within this test, therefore, the degradation is an extrapolation in galaxy number rather than a response to subgrid dimensions absent from the training set. Whether the same holds for summaries that carry more information than galaxy positions, velocities and number, such as projected maps or the matter power spectrum, we leave to future work.

\subsection{Implications}
\label{sec:gnn_discussion}

This failure is consistent with earlier tests of graph-based inference with galaxies, which found robustness to be the exception rather than the rule. \citet{desanti2023galaxies} trained networks to infer $\Omega_{\rm m}$ on the ASTRID, SIMBA, and IllustrisTNG suites in $(25~\Mpch)^3$ boxes: the ASTRID-trained network extrapolated well to the other suites, but the networks trained on SIMBA and IllustrisTNG were not robust across simulations. They associated the success of the ASTRID network mainly with the far broader range of galaxy numbers in its training catalogs, from about $30$ to $5000$ per box. CAMELS-CROCODILE extends this test to an independent feedback model and larger volumes, while confronting the IllustrisTNG-trained network with the same type of galaxy-number extrapolation: $23\%$ of the test graphs contain more galaxies than any of the $1024$ SB35 graphs ($2.4\times10^4$ at most), and nearly half exceed the SB35 90th percentile. Restricting the test to the graphs within the SB35 range reduces the $\Omega_{\rm m}$ error by more than a factor of two (Section~\ref{sec:gnn_results}).

The direction of the failure is itself informative. Because galaxy number increases with $\Omega_{\rm m}$ and CAMELS-CROCODILE produces more galaxies than IllustrisTNG, a network that simply read galaxy abundance as matter density would be expected to over-predict $\Omega_{\rm m}$ for the richest CROCODILE catalogs. The catastrophic outliers instead fall far below the truth, at $0.07$--$0.12$, and are essentially all among the runs whose galaxy numbers exceed the training range (open circles in Figure~\ref{fig:gnn}). The IllustrisTNG-trained network of \citet{desanti2023galaxies} failed in the same way on Magneticum: its residuals grow almost linearly with the true $\Omega_{\rm m}$, from nearly zero at $\Omega_{\rm m}=0.1$ to about $0.35$ at $0.5$ (mean bias $0.14$; their Figure~8), while its predictions remained nearly fixed at $\Omega_{\rm m}\simeq0.1$--$0.15$ across the Magneticum catalogs, including catalogs whose true values reached $\Omega_{\rm m}=0.5$. In both cases, graphs richer than any seen in training collapse the inference toward the low edge of the prior, which supports the extrapolation interpretation rather than contradicting it. 
The controlled tests of Section~\ref{sec:gnn_results} explain the direction. The network takes the logarithm of the number of galaxies as an explicit global input \citep{villanueva2022graphs,desanti2023galaxies}, and within the training range it already responds to additional galaxies at fixed cosmology by lowering the inferred $\Omega_{\rm m}$ and raising $\sigma_8$. The collapse of the richest graphs toward low $\Omega_{\rm m}$ is that same response carried far beyond the training range, where $\log N_{\rm gal}$ reaches $4.9$ against a training maximum of $4.4$. 
Consistent with this picture, a majority of the out-of-range runs are over-predicted in $\sigma_8$ (Figure~\ref{fig:gnn}). The same sensitivity plausibly underlies the overconfident $\sigma_8$ posteriors that persist within the training range: across the Sobol set, galaxy number varies strongly with the feedback parameters at fixed cosmology (Section~\ref{sec:gnn_setup}), and $\sigma_8$ is the parameter through which the network absorbs such changes.

Our result thus extends the conclusion of \citet{desanti2023galaxies} to $(50~\Mpch)^3$ volumes, to $\sigma_8$ as well as $\Omega_{\rm m}$, and to an independent feedback model, and it adds that the failure comes with overconfident rather than inflated posteriors, for $\sigma_8$ even within the training range in galaxy number, so that it would not reveal itself in an application to data. Training sets that span a deliberately wide range of galaxy formation outcomes, as ASTRID and CAMELS-CROCODILE Sobol set do, together with uncertainty estimates that are sensitive to distribution shift, are a prerequisite for trusting field-level posteriors from real surveys.

Caveats remain: we test one trained network and one summary statistic, galaxy graphs, at $z=0$ on the first \NSBrunsGNN{} Sobol runs and without survey selection, and the result concerns generalization rather than the constraining power of a network trained on CAMELS-CROCODILE itself.

\section{Outlook}
\label{sec:outlook}

The full CAMELS-CROCODILE dataset (snapshots, halo and subhalo catalogs, and precomputed summary statistics) will be released publicly through the CAMELS data infrastructure \citep{villaescusa-navarro2023datarelease}, following the formats of the existing suites so that the community's analysis tools apply directly.

The suite is designed to enable a broad scientific program, and Section~\ref{sec:gnn} provides its first result: a galaxy-graph network trained on IllustrisTNG fails on CAMELS-CROCODILE, and does so with overconfident posteriors. That result sets the immediate ML program. The same frozen-model test should be extended to the other summaries on which networks were trained for the second-generation suite, the matter power spectrum and projected field maps \citep{genel2026camels50}, to establish whether the failure is specific to galaxy graphs. The 1P set can then attribute it: because each Osaka parameter is varied in isolation, the parameters that drive the degradation can be identified directly, including the $17$ astrophysical parameters with no IllustrisTNG counterpart. Finally, networks trained on CAMELS-CROCODILE, and jointly on CAMELS-CROCODILE and the IllustrisTNG suites, will show whether training across physically distinct implementations restores robustness, extending the cross-suite experiments of \citet{villaescusa-navarro2022multifield} and \citet{ni2023camels} to $(50~\Mpch)^3$ volumes and to an independent feedback model. Completing the remaining L50 Sobol runs (from 201 runs presented here to the planned 256 runs) and extension to version-2 of CROCODILE with a jet radio-mode AGN feedback (Oku et al. 2026, in prep.) will further enlarge the data available for both testing and training.

Building on the established strengths of the Osaka model in the diffuse gas regime \citep{nagamine2021feedback,oku2024crocodile}, we will further use these simulations to study baryon acoustic oscillation shifts \citep{Sinigaglia24ApJL}, the Lyman-$\alpha$ forest \citep{Sinigaglia21,Sinigaglia22,Sinigaglia24AA,Nakashima25}, the metal enrichment of the CGM and IGM traced by metal-line absorbers (e.g., C\,\textsc{iv}, O\,\textsc{vi}) as a function of impact parameter \citep{Tumlinson17}, and cross-correlations with neutral hydrogen \citep{Ando19,Ando20,Murakami24MNRAS,Murakami24PRD} and fast-radio-burst dispersion measures \citep{crocodile2025frb}, where the model's metal-rich SN and AGN outflows make specific, testable predictions that differ qualitatively from the TNG-, SIMBA-, and ASTRID-family models. Together with the IllustrisTNG-based second-generation suite \citep{genel2026camels50}, CAMELS-CROCODILE broadens the foundation for simultaneously constraining baryonic and cosmological parameters from the non-linear Universe.

\begin{acknowledgments}
This project was supported by computational resources at the University of Osaka and the Simons Foundation. 
The authors thank the CAMELS collaboration for access to the simulation and analysis infrastructure. 
The Flatiron Institute is supported by the Simons Foundation. 
KN is grateful to the Flatiron Institute and the Simons Foundation for the travel support to visit CCA, during which part of this work was conducted. 
KN acknowledges support from JSPS KAKENHI grants 24H00002, 24H00241, JP25K01032. 
AJN is partly supported by JSPS KAKENHI grants JP25H01551, JP23H00108. 
KN, YO, and AJN are supported by the JSPS International Leading Research (ILR) project, JP22K21349. 
KN also acknowledges support from the Kavli IPMU, the World Premier Research Center Initiative (WPI), UTIAS, and the University of Tokyo.
DAA acknowledges support from NSF CAREER award AST-2442788, an Alfred P. Sloan Research Fellowship, and Cottrell Scholar Award CS-CSA-2023-028 by the Research Corporation for Science Advancement.
ChatGPT-5 (OpenAI 2025) and Claude (Anthropic 2026) were used to help format certain tables and figures and polish the language. All text generated by these tools was carefully reviewed, edited, and verified for accuracy by the authors, who assume full responsibility for the outcome.

\end{acknowledgments}

\bibliography{crocodile_camels}

\begin{thebibliography}{}
\expandafter\ifx\csname natexlab\endcsname\relax\def\natexlab#1{#1}\fi
\providecommand{\url}[1]{\href{#1}{#1}}
\providecommand{\dodoi}[1]{doi:~\href{http://doi.org/#1}{\nolinkurl{#1}}}
\providecommand{\doeprint}[1]{\href{http://ascl.net/#1}{\nolinkurl{http://ascl.net/#1}}}
\providecommand{\doarXiv}[1]{\href{https://arxiv.org/abs/#1}{\nolinkurl{https://arxiv.org/abs/#1}}}

\bibitem[{R. {Ando} {et~al.}(2019){Ando}, {Nishizawa}, {Hasegawa}, {Shimizu},
  \& {Nagamine}}]{Ando19}
{Ando}, R., {Nishizawa}, A.~J., {Hasegawa}, K., {Shimizu}, I., \& {Nagamine},
  K. 2019, \bibinfo{title}{{Redshift space distortion of 21 cm line at 1 < z <
  5 with cosmological hydrodynamic simulations},} \mnras, 484, 5389,
  \dodoi{10.1093/mnras/stz319}

\bibitem[{R. {Ando} {et~al.}(2021){Ando}, {Nishizawa}, {Shimizu}, \&
  {Nagamine}}]{Ando20}
{Ando}, R., {Nishizawa}, A.~J., {Shimizu}, I., \& {Nagamine}, K. 2021,
  \bibinfo{title}{{Reconstructing HI power spectrum with minimal parameters
  using the dark matter distribution beyond haloes},} \mnras, 507, 2937,
  \dodoi{10.1093/mnras/stab2284}

\bibitem[{S. {Aoyama} {et~al.}(2020){Aoyama}, {Hirashita}, \&
  {Nagamine}}]{Aoyama20}
{Aoyama}, S., {Hirashita}, H., \& {Nagamine}, K. 2020, \bibinfo{title}{{Galaxy
  simulation with the evolution of grain size distribution},} \mnras, 491,
  3844, \dodoi{10.1093/mnras/stz3253}

\bibitem[{S. {Aoyama} {et~al.}(2018){Aoyama}, {Hou}, {Hirashita}, {Nagamine},
  \& {Shimizu}}]{Aoyama18}
{Aoyama}, S., {Hou}, K.-C., {Hirashita}, H., {Nagamine}, K., \& {Shimizu}, I.
  2018, \bibinfo{title}{{Cosmological simulation with dust formation and
  destruction},} \mnras, 478, 4905, \dodoi{10.1093/mnras/sty1431}

\bibitem[{S. {Aoyama} {et~al.}(2019){Aoyama}, {Hirashita}, {Lim}, {Chang},
  {Wang}, {Nagamine}, {Hou}, {Shimizu}, {Chung}, {Lee}, \& {Zheng}}]{Aoyama19}
{Aoyama}, S., {Hirashita}, H., {Lim}, C.-F., {et~al.} 2019,
  \bibinfo{title}{{Comparison of cosmological simulations and deep
  submillimetre galaxy surveys},} \mnras, 484, 1852,
  \dodoi{10.1093/mnras/stz021}

\bibitem[{P.~S. {Behroozi} {et~al.}(2013){Behroozi}, {Wechsler}, \&
  {Conroy}}]{Behroozi13}
{Behroozi}, P.~S., {Wechsler}, R.~H., \& {Conroy}, C. 2013,
  \bibinfo{title}{{The Average Star Formation Histories of Galaxies in Dark
  Matter Halos from z = 0-8},} \apj, 770, 57,
  \dodoi{10.1088/0004-637X/770/1/57}

\bibitem[{S. {Bird}(2017){Bird}}]{Bird17fake}
{Bird}, S. 2017, \bibinfo{title}{{FSFE: Fake Spectra Flux Extractor},},
  Astrophysics Source Code Library, record ascl:1710.012

\bibitem[{G.~R. {Blumenthal} {et~al.}(1984){Blumenthal}, {Faber}, {Primack}, \&
  {Rees}}]{Blumenthal84}
{Blumenthal}, G.~R., {Faber}, S.~M., {Primack}, J.~R., \& {Rees}, M.~J. 1984,
  \bibinfo{title}{{Formation of galaxies and large-scale structure with cold
  dark matter},} \nat, 311, 517, \dodoi{10.1038/311517a0}

\bibitem[{Y. {Chen} {et~al.}(2020){Chen}, {Steidel}, {Hummels}, {Rudie},
  {Dong}, {Trainor}, {Bogosavljevi{\'c}}, {Erb}, {Pettini}, {Reddy}, {Shapley},
  {Strom}, {Theios}, {Faucher-Gigu{\`e}re}, {Hopkins}, \&
  {Kere{\v{s}}}}]{Chen20}
{Chen}, Y., {Steidel}, C.~C., {Hummels}, C.~B., {et~al.} 2020,
  \bibinfo{title}{{The Keck Baryonic Structure Survey: using
  foreground/background galaxy pairs to trace the structure and kinematics of
  circumgalactic neutral hydrogen at z 2},} \mnras, 499, 1721,
  \dodoi{10.1093/mnras/staa2808}

\bibitem[{R.~A. {Crain} {et~al.}(2015){Crain}, {Schaye}, {Bower}, {Furlong},
  {Schaller}, {Theuns}, {Dalla Vecchia}, {Frenk}, {McCarthy}, {Helly},
  {Jenkins}, {Rosas-Guevara}, {White}, \& {Trayford}}]{Crain15}
{Crain}, R.~A., {Schaye}, J., {Bower}, R.~G., {et~al.} 2015,
  \bibinfo{title}{{The EAGLE simulations of galaxy formation: calibration of
  subgrid physics and model variations},} \mnras, 450, 1937,
  \dodoi{10.1093/mnras/stv725}

\bibitem[{N.~S.~M. {de Santi} {et~al.}(2023){de Santi}, {Shao},
  {Villaescusa-Navarro}, {Abramo}, {Teyssier}, {Villanueva-Domingo},
  {et~al.}}]{desanti2023galaxies}
{de Santi}, N. S.~M., {Shao}, H., {Villaescusa-Navarro}, F., {et~al.} 2023,
  \bibinfo{title}{{Robust Field-level Likelihood-free Inference with
  Galaxies},} \apj, 952, 69, \dodoi{10.3847/1538-4357/acd1e2}

\bibitem[{A.~M. {Delgado} {et~al.}(2023){Delgado}, {Angl{\'e}s-Alc{\'a}zar},
  {Thiele}, {Pandey}, {Lehman}, {Somerville}, {Ntampaka}, {Genel},
  {Villaescusa-Navarro}, \& {Hernquist}}]{Delgado23}
{Delgado}, A.~M., {Angl{\'e}s-Alc{\'a}zar}, D., {Thiele}, L., {et~al.} 2023,
  \bibinfo{title}{{Predicting the impact of feedback on matter clustering with
  machine learning in CAMELS},} \mnras, 526, 5306,
  \dodoi{10.1093/mnras/stad2992}

\bibitem[{C.-A. {Faucher-Gigu{\`e}re} {et~al.}(2008){Faucher-Gigu{\`e}re},
  {Lidz}, {Hernquist}, \& {Zaldarriaga}}]{Faucher08a}
{Faucher-Gigu{\`e}re}, C.-A., {Lidz}, A., {Hernquist}, L., \& {Zaldarriaga}, M.
  2008, \bibinfo{title}{{Evolution of the Intergalactic Opacity: Implications
  for the Ionizing Background, Cosmic Star Formation, and Quasar Activity},}
  \apj, 688, 85, \dodoi{10.1086/592289}

\bibitem[{M. {Gebhardt} {et~al.}(2024){Gebhardt}, {Angl{\'e}s-Alc{\'a}zar},
  {Borrow}, {Genel}, {Villaescusa-Navarro}, {Ni}, {Lovell}, {Nagai},
  {Dav{\'e}}, {Marinacci}, {Vogelsberger}, \& {Hernquist}}]{Gebhardt24}
{Gebhardt}, M., {Angl{\'e}s-Alc{\'a}zar}, D., {Borrow}, J., {et~al.} 2024,
  \bibinfo{title}{{Cosmological baryon spread and impact on matter clustering
  in CAMELS},} \mnras, 529, 4896, \dodoi{10.1093/mnras/stae817}

\bibitem[{S. {Genel} {et~al.}(2026){Genel}, {Jo}, {Oh}, {Tillman}, {Lee},
  {Lee}, {Hern\'andez-Mart\'inez}, {Lovell}, {Sims}, {Burkhart}, {Nagamine},
  {Angl\'es-Alc\'azar}, \& {Villaescusa-Navarro}}]{genel2026camels50}
{Genel}, S., {Jo}, Y., {Oh}, B.~K., {et~al.} 2026, \bibinfo{title}{{Learning
  the Universe with the 2nd Generation of CAMELS: Varying 35 parameters of the
  IllustrisTNG model in (50 $h^{-1}$ Mpc)$^3$ boxes},} submitted to \apj.
\newblock \doarXiv{2606.10038}

\bibitem[{P. {Granizo} {et~al.}(2026){Granizo} {et~al.}}]{granizo2025agora}
{Granizo}, P., {et~al.} 2026, \bibinfo{title}{{Validating the CROCODILE Model
  Within the AGORA Galaxy Simulation Framework},} Galaxies, 14, 14,
  \dodoi{10.3390/galaxies14010014}

\bibitem[{J. {Greene} {et~al.}(2022){Greene}, {Bezanson}, {Ouchi}, {Silverman},
  \& {the PFS Galaxy Evolution Working Group}}]{Greene22}
{Greene}, J., {Bezanson}, R., {Ouchi}, M., {Silverman}, J., \& {the PFS Galaxy
  Evolution Working Group}. 2022, \bibinfo{title}{{The Prime Focus Spectrograph
  Galaxy Evolution Survey},} arXiv e-prints, arXiv:2206.14908.
\newblock \doarXiv{2206.14908}

\bibitem[{F. {Haardt} \& P. {Madau}(2012){Haardt} \& {Madau}}]{Haardt12}
{Haardt}, F., \& {Madau}, P. 2012, \bibinfo{title}{{Radiative Transfer in a
  Clumpy Universe. IV. New Synthesis Models of the Cosmic UV/X-Ray
  Background},} ApJ, 746, 125, \dodoi{10.1088/0004-637X/746/2/125}

\bibitem[{G. {Hinshaw} {et~al.}(2013){Hinshaw}, {Larson}, {Komatsu}, {Spergel},
  {Bennett}, {Dunkley}, {Nolta}, {Halpern}, {Hill}, {Odegard}, {Page}, {Smith},
  {Weiland}, {Gold}, {Jarosik}, {Kogut}, {Limon}, {Meyer}, {Tucker}, {Wollack},
  \& {Wright}}]{Hinshaw13}
{Hinshaw}, G., {Larson}, D., {Komatsu}, E., {et~al.} 2013,
  \bibinfo{title}{{Nine-year Wilkinson Microwave Anisotropy Probe (WMAP)
  Observations: Cosmological Parameter Results},} ApJS, 208, 19,
  \dodoi{10.1088/0067-0049/208/2/19}

\bibitem[{P.~F. {Hopkins}(2013){Hopkins}}]{Hopk13b}
{Hopkins}, P.~F. 2013, \bibinfo{title}{{A general class of Lagrangian smoothed
  particle hydrodynamics methods and implications for fluid mixing problems},}
  MNRAS, 428, 2840, \dodoi{10.1093/mnras/sts210}

\bibitem[{M. {Jung} {et~al.}(2025){Jung}, {Kim}, {Nguyen}, {Rodriguez-Cardoso},
  {Roca-F{\`a}brega}, {et~al.}}]{Jung25}
{Jung}, M., {Kim}, J.-h., {Nguyen}, T.~H., {et~al.} 2025, \bibinfo{title}{{The
  AGORA High-resolution Galaxy Simulations Comparison Project. VIII. Disk
  Formation and Evolution of Simulated Milky Way Mass Galaxy Progenitors at
  $1<z<5$},} \apj, 994, 245.
\newblock \doarXiv{2505.05720}

\bibitem[{H. {Kim} {et~al.}(2026){Kim}, {Kim}, {Jung}, {Roca-F{\`a}brega},
  {Ceverino}, {Granizo}, {Nagamine}, {et~al.}}]{Kim26}
{Kim}, H., {Kim}, J.-h., {Jung}, M., {et~al.} 2026, \bibinfo{title}{{The AGORA
  High-resolution Galaxy Simulations Comparison Project. X. Formation and
  Evolution of Galaxies at the High-redshift Frontier},} \apj, 1000, 276,
  \dodoi{10.3847/1538-4357/ae4a23}

\bibitem[{C.~C. {Lovell} {et~al.}(2025){Lovell}, {Starkenburg}, {Ho},
  {Angl{\'e}s-Alc{\'a}zar}, {Dav{\'e}}, {Gabrielpillai}, {Iyer}, {Matthews},
  {Roper}, {Somerville}, {Sommovigo}, \& {Villaescusa-Navarro}}]{Lovell25}
{Lovell}, C.~C., {Starkenburg}, T., {Ho}, M., {et~al.} 2025,
  \bibinfo{title}{{Learning the Universe: cosmological and astrophysical
  parameter inference with galaxy luminosity functions and colours},} \mnras,
  544, 3949, \dodoi{10.1093/mnras/staf1888}

\bibitem[{P. {Madau} \& M. {Dickinson}(2014){Madau} \&
  {Dickinson}}]{madau2014review}
{Madau}, P., \& {Dickinson}, M. 2014, \bibinfo{title}{{Cosmic Star-Formation
  History},} \araa, 52, 415, \dodoi{10.1146/annurev-astro-081811-125615}

\bibitem[{A. {Meiksin} {et~al.}(2017){Meiksin}, {Bolton}, \&
  {Puchwein}}]{Meiksin17}
{Meiksin}, A., {Bolton}, J.~S., \& {Puchwein}, E. 2017, \bibinfo{title}{{Gas
  around galaxy haloes - III: hydrogen absorption signatures around galaxies
  and QSOs in the Sherwood simulation suite},} \mnras, 468, 1893,
  \dodoi{10.1093/mnras/stx191}

\bibitem[{R. {Momose} {et~al.}(2021{\natexlab{a}}){Momose}, {Shimasaku},
  {Nagamine}, {Shimizu}, {Kashikawa}, {Ando}, \& {Kusakabe}}]{Momose21c}
{Momose}, R., {Shimasaku}, K., {Nagamine}, K., {et~al.} 2021{\natexlab{a}},
  \bibinfo{title}{{Catch Me if You Can: Biased Distribution of
  Ly{\ensuremath{\alpha}}-emitting Galaxies according to the Viewing
  Direction},} \apjl, 912, L24, \dodoi{10.3847/2041-8213/abf04c}

\bibitem[{R. {Momose} {et~al.}(2021{\natexlab{b}}){Momose}, {Shimizu},
  {Nagamine}, {Shimasaku}, {Kashikawa}, \& {Kusakabe}}]{Momose21a}
{Momose}, R., {Shimizu}, I., {Nagamine}, K., {et~al.} 2021{\natexlab{b}},
  \bibinfo{title}{{Connection between Galaxies and H I in Circumgalactic and
  Intergalactic Media: Variation according to Galaxy Stellar Mass and Star
  Formation Activity},} \apj, 911, 98, \dodoi{10.3847/1538-4357/abe1b9}

\bibitem[{B.~P. {Moster} {et~al.}(2013){Moster}, {Naab}, \& {White}}]{Moster13}
{Moster}, B.~P., {Naab}, T., \& {White}, S. D.~M. 2013,
  \bibinfo{title}{{Galactic star formation and accretion histories from
  matching galaxies to dark matter haloes},} \mnras, 428, 3121,
  \dodoi{10.1093/mnras/sts261}

\bibitem[{S. {Mukae} {et~al.}(2017){Mukae}, {Ouchi}, {Kakiichi}, {Suzuki},
  {Ono}, {Cai}, {Inoue}, {Chiang}, {Shibuya}, \& {Matsuda}}]{Mukae17}
{Mukae}, S., {Ouchi}, M., {Kakiichi}, K., {et~al.} 2017,
  \bibinfo{title}{{Cosmic Galaxy-IGM HI Relation at z {\tilde} 2-3 Probed in
  the COSMOS/UltraVISTA 1.6 Deg2 Field},} \apj, 835, 281,
  \dodoi{10.3847/1538-4357/835/2/281}

\bibitem[{K. {Murakami} {et~al.}(2024{\natexlab{a}}){Murakami}, {Nishizawa},
  {Nagamine}, \& {Shimizu}}]{Murakami24MNRAS}
{Murakami}, K., {Nishizawa}, A.~J., {Nagamine}, K., \& {Shimizu}, I.
  2024{\natexlab{a}}, \bibinfo{title}{{Impact of astrophysical effects on the
  dark matter mass constraint with 21 cm intensity mapping},} \mnras, 530,
  2052, \dodoi{10.1093/mnras/stae985}

\bibitem[{K. {Murakami} {et~al.}(2024{\natexlab{b}}){Murakami}, {Nishizawa},
  {Shimizu}, {Nagamine}, {et~al.}}]{Murakami24PRD}
{Murakami}, K., {Nishizawa}, A.~J., {Shimizu}, I., {Nagamine}, K., {et~al.}
  2024{\natexlab{b}}, \bibinfo{title}{{Constraining dark matter models using 21
  cm line intensity mapping},} Phys. Rev. D, 110, 023526,
  \dodoi{10.1103/PhysRevD.110.023526}

\bibitem[{K. {Nagamine} {et~al.}(2021){Nagamine}, {Shimizu}, {Fujita},
  {Suzuki}, {Lee}, {Momose}, {Mukae}, {Liang}, {Kashikawa}, {Ouchi}, \&
  {Silverman}}]{nagamine2021feedback}
{Nagamine}, K., {Shimizu}, I., {Fujita}, K., {et~al.} 2021,
  \bibinfo{title}{{Probing Feedback via IGM tomography and the
  Ly{\ensuremath{\alpha}} Forest with Subaru PFS, TMT/ELT, and JWST},} \apj,
  914, 66, \dodoi{10.3847/1538-4357/abfa16}

\bibitem[{K. {Nakashima} {et~al.}(2025){Nakashima}, {Nishizawa}, {Nagamine},
  {Shimizu}, \& {Oku}}]{Nakashima25}
{Nakashima}, K., {Nishizawa}, A.~J., {Nagamine}, K., {Shimizu}, I., \& {Oku},
  Y. 2025, \bibinfo{title}{{Lyman-$\alpha$ forest power spectrum and its
  cross-correlation with dark matter haloes in different astrophysical
  models},} \mnras, 537, 1343, \dodoi{10.1093/mnras/staf097}

\bibitem[{Y. {Ni} {et~al.}(2023){Ni}, {Genel}, {Angl\'es-Alc\'azar},
  {Villaescusa-Navarro}, {Jo}, {Bird}, {Di Matteo}, {Croft}, {Chen}, {de
  Santi}, {Gebhardt}, {Shao}, {Pandey}, {Hernquist}, \&
  {Dav\'e}}]{ni2023camels}
{Ni}, Y., {Genel}, S., {Angl\'es-Alc\'azar}, D., {et~al.} 2023,
  \bibinfo{title}{{The CAMELS Project: Expanding the Galaxy Formation Model
  Space with New ASTRID and 28-parameter TNG and SIMBA Suites},} \apj, 959,
  136, \dodoi{10.3847/1538-4357/ad022a}

\bibitem[{Y. {Oku} \& K. {Nagamine}(2024){Oku} \&
  {Nagamine}}]{oku2024crocodile}
{Oku}, Y., \& {Nagamine}, K. 2024, \bibinfo{title}{{Osaka Feedback Model. III.
  Cosmological Simulation CROCODILE},} \apj, 975, 183,
  \dodoi{10.3847/1538-4357/ad77d3}

\bibitem[{Y. {Oku} {et~al.}(2022){Oku}, {Tomida}, {Nagamine}, {Shimizu}, \&
  {Cen}}]{oku2022osaka}
{Oku}, Y., {Tomida}, K., {Nagamine}, K., {Shimizu}, I., \& {Cen}, R. 2022,
  \bibinfo{title}{{Osaka Feedback Model. II. Modeling Supernova Feedback Based
  on High-resolution Simulations},} \apjs, 262, 9,
  \dodoi{10.3847/1538-4365/ac77ff}

\bibitem[{{Planck Collaboration} {et~al.}(2020){Planck Collaboration},
  {Aghanim}, {Akrami}, {Ashdown}, {Aumont}, {Baccigalupi}, {Ballardini},
  {Banday}, {Barreiro}, {Bartolo}, {Basak}, {Battye}, {Benabed}, {Bernard},
  {Bersanelli}, {Bielewicz}, {Bock}, {Bond}, {Borrill}, {Bouchet}, {Boulanger},
  {Bucher}, {Burigana}, {Butler}, {Calabrese}, {Cardoso}, {Carron},
  {Challinor}, {Chiang}, {Chluba}, {Colombo}, {Combet}, {Contreras}, {Crill},
  {Cuttaia}, {de Bernardis}, {de Zotti}, {Delabrouille}, {Delouis}, {Di
  Valentino}, {Diego}, {Dor{\'e}}, {Douspis}, {Ducout}, {Dupac}, {Dusini},
  {Efstathiou}, {Elsner}, {En{\ss}lin}, {Eriksen}, {Fantaye}, {Farhang},
  {Fergusson}, {Fernandez-Cobos}, {Finelli}, {Forastieri}, {Frailis},
  {Fraisse}, {Franceschi}, {Frolov}, {Galeotta}, {Galli}, {Ganga},
  {G{\'e}nova-Santos}, {Gerbino}, {Ghosh}, {Gonz{\'a}lez-Nuevo}, {G{\'o}rski},
  {Gratton}, {Gruppuso}, {Gudmundsson}, {Hamann}, {Handley}, {Hansen},
  {Herranz}, {Hildebrandt}, {Hivon}, {Huang}, {Jaffe}, {Jones}, {Karakci},
  {Keih{\"a}nen}, {Keskitalo}, {Kiiveri}, {Kim}, {Kisner}, {Knox},
  {Krachmalnicoff}, {Kunz}, {Kurki-Suonio}, {Lagache}, {Lamarre}, {Lasenby},
  {Lattanzi}, {Lawrence}, {Le Jeune}, {Lemos}, {Lesgourgues}, {Levrier},
  {Lewis}, {Liguori}, {Lilje}, {Lilley}, {Lindholm}, {L{\'o}pez-Caniego},
  {Lubin}, {Ma}, {Mac{\'\i}as-P{\'e}rez}, {Maggio}, {Maino}, {Mandolesi},
  {Mangilli}, {Marcos-Caballero}, {Maris}, {Martin}, {Martinelli},
  {Mart{\'\i}nez-Gonz{\'a}lez}, {Matarrese}, {Mauri}, {McEwen}, {Meinhold},
  {Melchiorri}, {Mennella}, {Migliaccio}, {Millea}, {Mitra},
  {Miville-Desch{\^e}nes}, {Molinari}, {Montier}, {Morgante}, {Moss}, {Natoli},
  {N{\o}rgaard-Nielsen}, {Pagano}, {Paoletti}, {Partridge}, {Patanchon},
  {Peiris}, {Perrotta}, {Pettorino}, {Piacentini}, {Polastri}, {Polenta},
  {Puget}, {Rachen}, {Reinecke}, {Remazeilles}, {Renzi}, {Rocha}, {Rosset},
  {Roudier}, {Rubi{\~n}o-Mart{\'\i}n}, {Ruiz-Granados}, {Salvati}, {Sandri},
  {Savelainen}, {Scott}, {Shellard}, {Sirignano}, {Sirri}, {Spencer},
  {Sunyaev}, {Suur-Uski}, {Tauber}, {Tavagnacco}, {Tenti}, {Toffolatti},
  {Tomasi}, {Trombetti}, {Valenziano}, {Valiviita}, {Van Tent}, {Vibert},
  {Vielva}, {Villa}, {Vittorio}, {Wandelt}, {Wehus}, {White}, {White},
  {Zacchei}, \& {Zonca}}]{planck2020}
{Planck Collaboration}, {Aghanim}, N., {Akrami}, Y., {et~al.} 2020,
  \bibinfo{title}{{Planck 2018 results. VI. Cosmological parameters},} \aap,
  641, A6, \dodoi{10.1051/0004-6361/201833910}

\bibitem[{O. {Rakic} {et~al.}(2012){Rakic}, {Schaye}, {Steidel}, \&
  {Rudie}}]{Rakic12}
{Rakic}, O., {Schaye}, J., {Steidel}, C.~C., \& {Rudie}, G.~C. 2012,
  \bibinfo{title}{{Neutral Hydrogen Optical Depth near Star-forming Galaxies at
  z {\ensuremath{\approx}} 2.4 in the Keck Baryonic Structure Survey},} \apj,
  751, 94, \dodoi{10.1088/0004-637X/751/2/94}

\bibitem[{L.~E.~C. {Romano} {et~al.}(2022){Romano}, {Nagamine}, \&
  {Hirashita}}]{Romano22a}
{Romano}, L. E.~C., {Nagamine}, K., \& {Hirashita}, H. 2022,
  \bibinfo{title}{{Dust diffusion in SPH simulations of an isolated galaxy},}
  \mnras, 514, 1441, \dodoi{10.1093/mnras/stac1385}

\bibitem[{Y.~M. {Rosas-Guevara} {et~al.}(2015){Rosas-Guevara}, {Bower},
  {Schaye}, {Furlong}, {Frenk}, {Booth}, {Crain}, {Dalla Vecchia}, {Schaller},
  \& {Theuns}}]{rosas-guevara2015accretion}
{Rosas-Guevara}, Y.~M., {Bower}, R.~G., {Schaye}, J., {et~al.} 2015,
  \bibinfo{title}{{The impact of angular momentum on black hole accretion rates
  in simulations of galaxy formation},} \mnras, 454, 1038,
  \dodoi{10.1093/mnras/stv2056}

\bibitem[{T.~R. {Saitoh}(2017){Saitoh}}]{saitoh2017celib}
{Saitoh}, T.~R. 2017, \bibinfo{title}{{Chemical Evolution Library for Galaxy
  Formation Simulation},} \aj, 153, 85, \dodoi{10.3847/1538-3881/153/2/85}

\bibitem[{J. {Schaye} {et~al.}(2015){Schaye}, {Crain}, {Bower}, {Furlong},
  {Schaller}, {Theuns}, {Dalla Vecchia}, {Frenk}, {McCarthy}, {Helly},
  {Jenkins}, {Rosas-Guevara}, {White}, {Baes}, {Booth}, {Camps}, {Navarro},
  {Qu}, {Rahmati}, {Sawala}, {Thomas}, \& {Trayford}}]{Schaye15}
{Schaye}, J., {Crain}, R.~A., {Bower}, R.~G., {et~al.} 2015,
  \bibinfo{title}{{The EAGLE project: simulating the evolution and assembly of
  galaxies and their environments},} MNRAS, 446, 521,
  \dodoi{10.1093/mnras/stu2058}

\bibitem[{H. {Shao} {et~al.}(2023){Shao}, {Villaescusa-Navarro},
  {Villanueva-Domingo}, {Teyssier}, {Zhang}, {et~al.}}]{shao2023halos}
{Shao}, H., {Villaescusa-Navarro}, F., {Villanueva-Domingo}, P., {et~al.} 2023,
  \bibinfo{title}{{Robust Field-level Inference of Cosmological Parameters with
  Dark Matter Halos},} \apj, 944, 27, \dodoi{10.3847/1538-4357/acac7a}

\bibitem[{R.~K. Sheth \& G. Tormen(1999)Sheth \& Tormen}]{Sheth99}
Sheth, R.~K., \& Tormen, G. 1999, \bibinfo{title}{Large-scale bias and the peak
  background split,} MNRAS, 308, 119

\bibitem[{R.~K. Sheth \& G. Tormen(2002)Sheth \& Tormen}]{Sheth02}
Sheth, R.~K., \& Tormen, G. 2002, \bibinfo{title}{An excursion set model of
  hierarchical clustering: ellipsoidal collapse and the moving barrier,} MNRAS,
  329, 61

\bibitem[{I. {Shimizu} {et~al.}(2019){Shimizu}, {Todoroki}, {Yajima}, \&
  {Nagamine}}]{shimizu2019osaka}
{Shimizu}, I., {Todoroki}, K., {Yajima}, H., \& {Nagamine}, K. 2019,
  \bibinfo{title}{{Osaka feedback model: isolated disc galaxy simulations},}
  \mnras, 484, 2632, \dodoi{10.1093/mnras/stz098}

\bibitem[{F. {Sinigaglia} {et~al.}(2021){Sinigaglia}, {Kitaura},
  {Balaguera-Antol{\'\i}nez}, {Nagamine}, {Ata}, {Shimizu}, \&
  {S{\'a}nchez-Benavente}}]{Sinigaglia21}
{Sinigaglia}, F., {Kitaura}, F.-S., {Balaguera-Antol{\'\i}nez}, A., {et~al.}
  2021, \bibinfo{title}{{The Bias from Hydrodynamic Simulations: Mapping Baryon
  Physics onto Dark Matter Fields},} \apj, 921, 66,
  \dodoi{10.3847/1538-4357/ac158b}

\bibitem[{F. {Sinigaglia} {et~al.}(2022){Sinigaglia}, {Kitaura},
  {Balaguera-Antol{\'\i}nez}, {Shimizu}, {Nagamine}, {S{\'a}nchez-Benavente},
  \& {Ata}}]{Sinigaglia22}
{Sinigaglia}, F., {Kitaura}, F.-S., {Balaguera-Antol{\'\i}nez}, A., {et~al.}
  2022, \bibinfo{title}{{Mapping the Three-dimensional Ly{\ensuremath{\alpha}}
  Forest Large-scale Structure in Real and Redshift Space},} \apj, 927, 230,
  \dodoi{10.3847/1538-4357/ac5112}

\bibitem[{F. {Sinigaglia} {et~al.}(2024{\natexlab{a}}){Sinigaglia}, {Kitaura},
  {Nagamine}, \& {Oku}}]{Sinigaglia24ApJL}
{Sinigaglia}, F., {Kitaura}, F.-S., {Nagamine}, K., \& {Oku}, Y.
  2024{\natexlab{a}}, \bibinfo{title}{{The Negative BAO Shift in the Ly$\alpha$
  Forest from Cosmological Simulations},} \apjl, 971, L22,
  \dodoi{10.3847/2041-8213/ad58e1}

\bibitem[{F. {Sinigaglia} {et~al.}(2024{\natexlab{b}}){Sinigaglia}, {Kitaura},
  {Nagamine}, {Oku}, \& {Balaguera-Antol{\'\i}nez}}]{Sinigaglia24AA}
{Sinigaglia}, F., {Kitaura}, F.-S., {Nagamine}, K., {Oku}, Y., \&
  {Balaguera-Antol{\'\i}nez}, A. 2024{\natexlab{b}},
  \bibinfo{title}{{Field-level {Ly$\alpha$} forest modelling in redshift space
  via augmented nonlocal Fluctuating Gunn-Peterson Approximation},} \aap, 682,
  A21, \dodoi{10.1051/0004-6361/202347152}

\bibitem[{D. {Sorini} {et~al.}(2024){Sorini}, {Bose}, {Dav\'e}, \&
  {Angl\'es-Alc\'azar}}]{sorini2024profiles}
{Sorini}, D., {Bose}, S., {Dav\'e}, R., \& {Angl\'es-Alc\'azar}, D. 2024,
  \bibinfo{title}{{The impact of feedback on the evolution of gas density
  profiles from galaxies to clusters: a universal fitting formula from the
  Simba suite of simulations},} The Open Journal of Astrophysics, 7, 115,
  \dodoi{10.33232/001c.126621}

\bibitem[{D. {Sorini} {et~al.}(2020){Sorini}, {Dav\'e}, \&
  {Angl\'es-Alc\'azar}}]{sorini2020simba}
{Sorini}, D., {Dav\'e}, R., \& {Angl\'es-Alc\'azar}, D. 2020,
  \bibinfo{title}{{SIMBA: the average properties of the circumgalactic medium
  of $2 \leq z \leq 3$ quasars are determined primarily by stellar feedback},}
  \mnras, 499, 2760, \dodoi{10.1093/mnras/staa2937}

\bibitem[{D. {Sorini} {et~al.}(2022){Sorini}, {Dav\'e}, {Cui}, \&
  {Appleby}}]{sorini2022baryons}
{Sorini}, D., {Dav\'e}, R., {Cui}, W., \& {Appleby}, S. 2022,
  \bibinfo{title}{{How baryons affect haloes and large-scale structure: a
  unified picture from the Simba simulation},} \mnras, 516, 883,
  \dodoi{10.1093/mnras/stac2214}

\bibitem[{D. {Sorini} \& J.~A. {Peacock}(2021){Sorini} \&
  {Peacock}}]{sorini2021sfr}
{Sorini}, D., \& {Peacock}, J.~A. 2021, \bibinfo{title}{{Extended
  Hernquist-Springel formalism for cosmic star formation},} \mnras, 508, 5802

\bibitem[{V. {Springel} \& L. {Hernquist}(2003){Springel} \&
  {Hernquist}}]{springel2003sfr}
{Springel}, V., \& {Hernquist}, L. 2003, \bibinfo{title}{{Cosmological smoothed
  particle hydrodynamics simulations: a hybrid multiphase model for star
  formation},} MNRAS, 339, 289, \dodoi{10.1046/j.1365-8711.2003.06206.x}

\bibitem[{V. {Springel} {et~al.}(2021){Springel}, {Pakmor}, {Zier}, \&
  {Reinecke}}]{Springel21}
{Springel}, V., {Pakmor}, R., {Zier}, O., \& {Reinecke}, M. 2021,
  \bibinfo{title}{{Simulating cosmic structure formation with the GADGET-4
  code},} \mnras, 506, 2871, \dodoi{10.1093/mnras/stab1855}

\bibitem[{C.~C. {Steidel} {et~al.}(2010){Steidel}, {Erb}, {Shapley}, {Pettini},
  {Reddy}, {Bogosavljevi{\'c}}, {Rudie}, \& {Rakic}}]{Steidel10}
{Steidel}, C.~C., {Erb}, D.~K., {Shapley}, A.~E., {et~al.} 2010,
  \bibinfo{title}{{The Structure and Kinematics of the Circumgalactic Medium
  from Far-ultraviolet Spectra of z \raisebox{-0.5ex}\textasciitilde= 2-3
  Galaxies},} \apj, 717, 289, \dodoi{10.1088/0004-637X/717/1/289}

\bibitem[{M.~T. {Tillman} {et~al.}(2025){Tillman}, {Burkhart}, {Tonnesen},
  {Bird}, \& {Bryan}}]{Tillman25}
{Tillman}, M.~T., {Burkhart}, B., {Tonnesen}, S., {Bird}, S., \& {Bryan}, G.~L.
  2025, \bibinfo{title}{{The Effects of Active Galactic Nuclei Feedback on the
  Ly$\alpha$ Forest Flux Power Spectrum},} \apj, 980, 72,
  \dodoi{10.3847/1538-4357/ada5f7}

\bibitem[{M.~T. {Tillman} {et~al.}(2023{\natexlab{a}}){Tillman}, {Burkhart},
  {Tonnesen}, {Bird}, {Bryan}, {Angl{\'e}s-Alc{\'a}zar}, {Dav{\'e}}, \&
  {Genel}}]{Tillman23a}
{Tillman}, M.~T., {Burkhart}, B., {Tonnesen}, S., {et~al.} 2023{\natexlab{a}},
  \bibinfo{title}{{Efficient Long-range AGN Feedback Affects the Low-redshift
  Ly$\alpha$ Forest},} \apjl, 945, L17, \dodoi{10.3847/2041-8213/acb7f1}

\bibitem[{M.~T. {Tillman} {et~al.}(2023{\natexlab{b}}){Tillman}, {Burkhart},
  {Tonnesen}, {Bird}, {Bryan}, {Angl{\'e}s-Alc{\'a}zar}, {Hassan},
  {Somerville}, {Dav{\'e}}, {Marinacci}, {Hernquist}, \&
  {Vogelsberger}}]{Tillman23}
{Tillman}, M.~T., {Burkhart}, B., {Tonnesen}, S., {et~al.} 2023{\natexlab{b}},
  \bibinfo{title}{{An Exploration of AGN and Stellar Feedback Effects in the
  Intergalactic Medium via the Low-redshift Ly$\alpha$ Forest},} \aj, 166, 228,
  \dodoi{10.3847/1538-3881/ad02f5}

\bibitem[{K. {Tomaru} {et~al.}(2026){Tomaru}, {Oku}, {Toyouchi}, \&
  {Nagamine}}]{Tomaru26}
{Tomaru}, K., {Oku}, Y., {Toyouchi}, D., \& {Nagamine}, K. 2026,
  \bibinfo{title}{{CROCODILE-DWARF: Assembly and Kinematics of Field Dwarf
  Galaxies with GADGET4-OSAKA},} \apj, 1002, 5.
\newblock \doarXiv{2510.26513}

\bibitem[{J. {Tumlinson} {et~al.}(2017){Tumlinson}, {Peeples}, \&
  {Werk}}]{Tumlinson17}
{Tumlinson}, J., {Peeples}, M.~S., \& {Werk}, J.~K. 2017, \bibinfo{title}{{The
  Circumgalactic Medium},} \araa, 55, 389,
  \dodoi{10.1146/annurev-astro-091916-055240}

\bibitem[{M.~L. {Turner} {et~al.}(2014){Turner}, {Schaye}, {Steidel}, {Rudie},
  \& {Strom}}]{Turner14}
{Turner}, M.~L., {Schaye}, J., {Steidel}, C.~C., {Rudie}, G.~C., \& {Strom},
  A.~L. 2014, \bibinfo{title}{{Metal-line absorption around z
  {\ensuremath{\approx}} 2.4 star-forming galaxies in the Keck Baryonic
  Structure Survey},} \mnras, 445, 794, \dodoi{10.1093/mnras/stu1801}

\bibitem[{M.~P. {van Daalen} {et~al.}(2020){van Daalen}, {McCarthy}, \&
  {Schaye}}]{vanDaalen20}
{van Daalen}, M.~P., {McCarthy}, I.~G., \& {Schaye}, J. 2020,
  \bibinfo{title}{{Exploring the effects of galaxy formation on matter
  clustering through a library of simulation power spectra},} \mnras, 491,
  2424, \dodoi{10.1093/mnras/stz3199}

\bibitem[{M.~L. {van Loon} \& M.~P. {van Daalen}(2024){van Loon} \& {van
  Daalen}}]{Loon24}
{van Loon}, M.~L., \& {van Daalen}, M.~P. 2024, \bibinfo{title}{{The
  contribution of massive haloes to the matter power spectrum in the presence
  of AGN feedback},} \mnras, 528, 4623, \dodoi{10.1093/mnras/stae285}

\bibitem[{F. {Villaescusa-Navarro} {et~al.}(2021){Villaescusa-Navarro},
  {Angl\'es-Alc\'azar}, {Genel}, {et~al.}}]{villaescusa-navarro2021camels}
{Villaescusa-Navarro}, F., {Angl\'es-Alc\'azar}, D., {Genel}, S., {et~al.}
  2021, \bibinfo{title}{{The CAMELS Project: Cosmology and Astrophysics with
  Machine-learning Simulations},} \apj, 915, 71,
  \dodoi{10.3847/1538-4357/abf7ba}

\bibitem[{F. {Villaescusa-Navarro} {et~al.}(2022){Villaescusa-Navarro}
  {et~al.}}]{villaescusa-navarro2022multifield}
{Villaescusa-Navarro}, F., {et~al.} 2022, \bibinfo{title}{{The CAMELS
  Multifield Data Set: Learning the Universe's Fundamental Parameters with
  Artificial Intelligence},} \apjs, 259, 61, \dodoi{10.3847/1538-4365/ac5ab0}

\bibitem[{F. {Villaescusa-Navarro} {et~al.}(2023){Villaescusa-Navarro}
  {et~al.}}]{villaescusa-navarro2023datarelease}
{Villaescusa-Navarro}, F., {et~al.} 2023, \bibinfo{title}{{The CAMELS Project:
  Public Data Release},} \apjs, 265, 54, \dodoi{10.3847/1538-4365/acbf47}

\bibitem[{P. {Villanueva-Domingo} \& F.
  {Villaescusa-Navarro}(2022){Villanueva-Domingo} \&
  {Villaescusa-Navarro}}]{villanueva2022graphs}
{Villanueva-Domingo}, P., \& {Villaescusa-Navarro}, F. 2022,
  \bibinfo{title}{{Learning Cosmology and Clustering with Cosmic Graphs},}
  \apj, 937, 115, \dodoi{10.3847/1538-4357/ac8930}

\bibitem[{Z. {Zhang} {et~al.}(2025){Zhang}, {Oku}, {Nagamine},
  {et~al.}}]{crocodile2025frb}
{Zhang}, Z., {Oku}, Y., {Nagamine}, K., {et~al.} 2025, \bibinfo{title}{{Probing
  the Cosmic Baryon Distribution and the Impact of Active Galactic Nuclei
  Feedback with Fast Radio Bursts in the CROCODILE Simulation},} \apj,
  \dodoi{10.3847/1538-4357/ae00c2}

\end{thebibliography}

\end{document}